\documentclass[twocolumn]{aastex62} 
\usepackage{amsmath}
\usepackage{amssymb}
\usepackage{mathtools}
\usepackage{mathrsfs}
\usepackage{bm}
\usepackage{enumerate}
\usepackage{booktabs}
\graphicspath{{./}{figures/}}

\newcommand{\project}[1]{\textsl{#1}}
\usepackage{xspace}

\newcommand*{\NICER}{\project{NICER}\xspace}
\newcommand*{\MultiNest}{\textsc{MultiNest}\xspace}

\newcommand{\msol}{M$_\odot$~}

\usepackage{hyperref}

\accepted{23/03/2020}

\submitjournal{ApJ Letters}

\shorttitle{Multi-messenger constraints on the EOS}
\shortauthors{Raaijmakers et al.}

\begin{document}

\correspondingauthor{G.~Raaijmakers}
\email{G.Raaijmakers@uva.nl}

\author[0000-0002-9397-786X]{G.~Raaijmakers}
\affil{GRAPPA, Anton Pannekoek Institute for Astronomy and Institute of High-Energy Physics, University of Amsterdam, Science Park 904, 1098 XH Amsterdam, Netherlands}

\author{S.~K.~Greif}
\affil{Institut f\"ur Kernphysik, Technische Universit\"at Darmstadt, 64289 Darmstadt, Germany}
\affil{ExtreMe Matter Institute EMMI, GSI Helmholtzzentrum f\"ur Schwerionenforschung GmbH, 64291 Darmstadt, Germany}

\author[0000-0001-9313-0493]{T.~E.~Riley}
\affil{Anton Pannekoek Institute for Astronomy, University of Amsterdam, Science Park 904, 1090GE Amsterdam, the Netherlands}

\author[0000-0002-3394-6105]{T.~Hinderer}
\affil{GRAPPA, Anton Pannekoek Institute for Astronomy and Institute of High-Energy Physics, University of Amsterdam, Science Park 904, 1098 XH Amsterdam, Netherlands}
\affil{Delta Institute for Theoretical Physics, Science Park 904, 1090 GL Amsterdam, The Netherlands}

\author{K.~Hebeler}
\affil{Institut f\"ur Kernphysik, Technische Universit\"at Darmstadt, 64289 Darmstadt, Germany}
\affil{ExtreMe Matter Institute EMMI, GSI Helmholtzzentrum f\"ur Schwerionenforschung GmbH, 64291 Darmstadt, Germany}

\author{A.~Schwenk}
\affil{Institut f\"ur Kernphysik, Technische Universit\"at Darmstadt, 64289 Darmstadt, Germany}
\affil{ExtreMe Matter Institute EMMI, GSI Helmholtzzentrum f\"ur Schwerionenforschung GmbH, 64291 Darmstadt, Germany}
\affil{Max-Planck-Institut f\"ur Kernphysik, Saupfercheckweg 1, 69117 Heidelberg, Germany}

\author[0000-0002-1009-2354]{A.~L.~Watts}
\affil{Anton Pannekoek Institute for Astronomy, University of Amsterdam, Science Park 904, 1090GE Amsterdam, the Netherlands}

\author[0000-0001-6573-7773]{S.~Nissanke}
\affil{GRAPPA, Anton Pannekoek Institute for Astronomy and Institute of High-Energy Physics, University of Amsterdam, Science Park 904, 1098 XH Amsterdam, Netherlands}
\affil{Nikhef, Science Park 105, 1098 XG Amsterdam, The Netherlands}

\author[0000-0002-6449-106X]{S.~Guillot}
\affil{IRAP, CNRS, 9 avenue du Colonel Roche, BP 44346, F-31028 Toulouse Cedex 4, France}
\affil{Universit\'{e} de Toulouse, CNES, UPS-OMP, F-31028 Toulouse, France.}

\author{J.~M.~Lattimer}
\affil{Department of Physics and Astronomy, Stony Brook University, Stony Brook, NY 11794-3800, USA}

\author[0000-0002-8961-939X]{R.~M.~Ludlam}
\affil{Cahill Center for Astronomy and Astrophysics, California Institute of Technology, Pasadena, CA 91125, USA}

\title{Constraining the dense matter equation of state with joint analysis of \NICER and LIGO/Virgo measurements}

\begin{abstract}
The \NICER collaboration recently published a joint estimate of the mass and the radius of PSR J0030+0451, derived via X-ray pulse-profile modeling. \citet{Raaijmakers19} explored the implications of this measurement for the dense matter equation of state (EOS) using two parameterizations of the high-density EOS: a piecewise-polytropic model, and a model based on the speed of sound in neutron stars. In this work we obtain further constraints on the EOS following this approach, but we also include information about the tidal deformability of neutron stars from the gravitational wave signal of the compact binary merger GW170817. We compare the constraints on the EOS to those set by the recent measurement of a $2.14$ M$_{\odot}$ pulsar, included as a likelihood function approximated by a Gaussian, and find a small increase in information gain. To show the flexibility of our method, we also explore the possibility that GW170817 was a neutron star-black hole merger, which yields weaker constraints on the EOS. 
\end{abstract}


\section{Introduction} \label{sec:intro}
Determining the behavior of matter at supranuclear densities is one of the major challenges of modern astrophysics and nuclear physics. Astronomical multi-messenger observations yield statistical measurements of neutron star (NS) properties such as gravitational mass, radius, and tidal deformability, providing a way to study matter under extreme conditions. Theoretical predictions for the phases of matter in NS cores span a wide range, from neutron-rich nucleonic matter to hyperon or deconfined quark formation or the emergence of a Bose-Einstein condensate or a color superconducting phase \citep[see][for recent reviews on this topic]{Hebeler15,Lattimer16,Oertel17,Baym18}. In practice, our uncertainty about dense matter is usually expressed in terms of a space of viable equation of state (EOS) models \citep[see, e.g.,][and references therein]{Abbott18,Raaijmakers19}.

Recently NASA's \textit{Neutron Star Interior Composition Explorer} (\NICER), an X-ray telescope on board the \textit{International Space Station}, has delivered a joint mass-radius measurement for the millisecond pulsar (MSP) PSR~J0030$+$0451 using pulse-profile modeling \citep[see][and references therein for a description of the technique]{Watts19b}. Two independent analyses were conducted within the collaboration, each making slightly different assumptions about the modeling \citep[including priors;][]{Miller19, Riley19}. The results depended strongly on the assumed geometry for the X-ray-emitting surface hot regions, but it was possible to identify a superior configuration based principally on the likelihood. The results of the two analyses were, however, deemed consistent: \citet{Riley19} reported an inferred mass and equatorial radius of $M=1.34^{+0.15}_{-0.16}$~\msol and $R_\mathrm{eq} = 12.71^{+1.14}_{-1.19}$~km (for the 68\% credible interval); \citet{Miller19}, on the other hand, found $M=1.44^{+0.15}_{-0.14}$~\msol and $R_\mathrm{eq} = 13.02^{+1.24}_{-1.06}$~km.

Constraints on the mass and radius have recently also been obtained from the gravitational wave (GW) observations of the binary NS merger event GW170817~\citep{GW170817discovery}. The LIGO Scientific and Virgo Collaborations (LVC) reported measurements of the masses and EOS-dependent tidal deformability parameters of the NSs under different prior assumptions on the spins and using various waveform models~\citep{GW170817discovery,GW170817sourceproperties,LIGO_161218_catalog}; see \citet{Kastaun:2019bxo} for a critical re-examination of the results. The corresponding radii and EOS constraints were inferred in two ways, by using a parameterized spectral EOS~\citep{Lindblom10}
and by employing EOS-insensitive relations, both using the low-spin priors ($cS/GM^2<0.05$ where $S$ is the spin angular momentum) and hence non-rotating stellar models. The results were consistent, leading to $R=11.9^{+1.4}_{-1.4}$~km from the spectral EOS analysis and $R=10.8^{+2.0}_{-1.7}$~km for the more massive NS at the $90\%$ credible interval~\citep{Abbott18}; see also \citet{De18,Annala18,Tews:2018chv} for independent related work. A number of studies have further included additional constraints from the electromagnetic counterparts assuming a NS-NS progenitor~\citep{Bauswein:2017vtn,Gao:2017fcu,Coughlin:2018miv,Most:2018hfd,Capano:2019eae,Margalit:2019dpi,Radice:2018ozg,Shibata:2019ctb}. In this work, we remain cautious of the large uncertainties in modeling the electromagnetic counterparts and only use the fact that the observed kilonova, an ultraviolet-optical-infrared transient powered by rapid neutron-capture nucleosynthesis \citep[see, e.g., ][]{1976ApJ...210..549L,1998ApJ...507L..59L,1999A&A...341..499R, 2005astro.ph.10256K,2010MNRAS.406.2650M,Metzger2017} indicated that the progenitor binary involved at least one NS. We consider two possibilities, a double NS system as assumed in most analyses and a NS-black hole binary. The mass of the black hole (BH) in the latter scenario would be very low and could have originated from an earlier merger of two NSs or be of primordial origin \citep[see, e.g.,][]{Yang:2017gfb, Coughlin19, Hinderer19}.

 In this Letter, we perform a joint analysis of the EOS constraints from \NICER and GW170817, following the method for connecting global NS parameters to the EOS used in~\citet{Raaijmakers19}.  We focus on the mass-radius measurement from \citet{Riley19}\footnote{The posterior samples from that analysis are available in a Zenodo repository \citet{samples}. We have now updated the repository from the version published alongside \citet{Riley19} to include the following additional material:  files containing only the mass-radius samples, contour files for the credible regions in mass-radius space, and simplified coordinate files for the boundaries of the hot regions on the stellar surface.}, as the measurement of \citet{Miller19} has already been used to jointly constrain the EOS with GW and radio pulsar measurements. We will compare our findings to those of \citet{Miller19} in Section~\ref{sec:disc}.
 
 We use EOS models that incorporate prior information from nuclear physics up to around saturation density, and two different parameterized extensions at high density; one using piecewise-polytropes, and one based on physically motivated assumptions about the speed of sound. We develop the methodology for the combined interpretation of these measurements in a Bayesian framework that also takes into account the measurements of massive pulsars. Our method can readily include a larger number of NSs from anticipated future multi-messenger observations. Next, we analyze the impact of systematic uncertainties arising from different priors for the EOS. We show that the priors used for the spectral EOS inference from GWs~\citep{Abbott18} allow for much stiffer EOSs than the priors in ~\citet{Raaijmakers19} and explain the reasons for these differences. Nevertheless, we find that the resulting EOS constraints are broadly consistent. We quantify explicitly that the fact that GW measurements determine the chirp mass to high accuracy can be utilized to accelerate the parameter inference by treating it as fixed.   
 
\begin{figure}[t!]
\centering
\includegraphics[width=.48\textwidth]{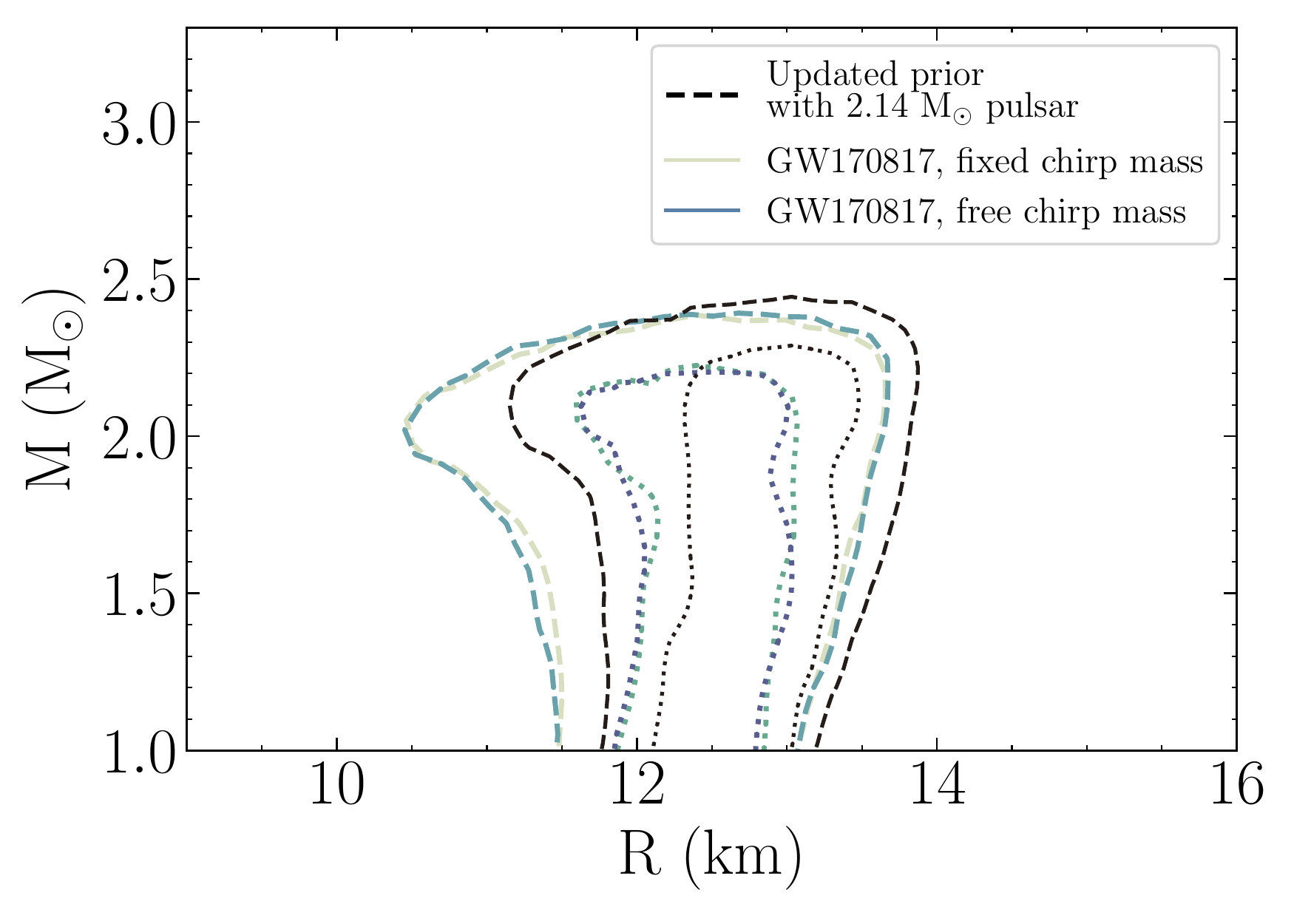}
\caption{Comparison of the prior on EOS parameters, transformed to the space of mass and radius, as it would be updated after performing parameter estimation on GW170817 with (green contours) and without (blue contours) fixing the chirp mass to the median, $M_{\text{chirp}} = 1.186$~M$_{\odot}$, of its marginal posterior distribution. The two distributions show some small-scale differences in the $1\sigma$ contour but are globally consistent. For comparison we also show the prior distribution before including information from GW170817, but with the $2.14$ M$_{\odot}$ pulsar information, in black contours. For all contours the dotted and dashed lines indicate the $68\%$ and $95\%$ credible regions, respectively.}
\label{fig:fig1}
\end{figure}

\begin{figure*}[t!]
\centering
\includegraphics[width=\textwidth]{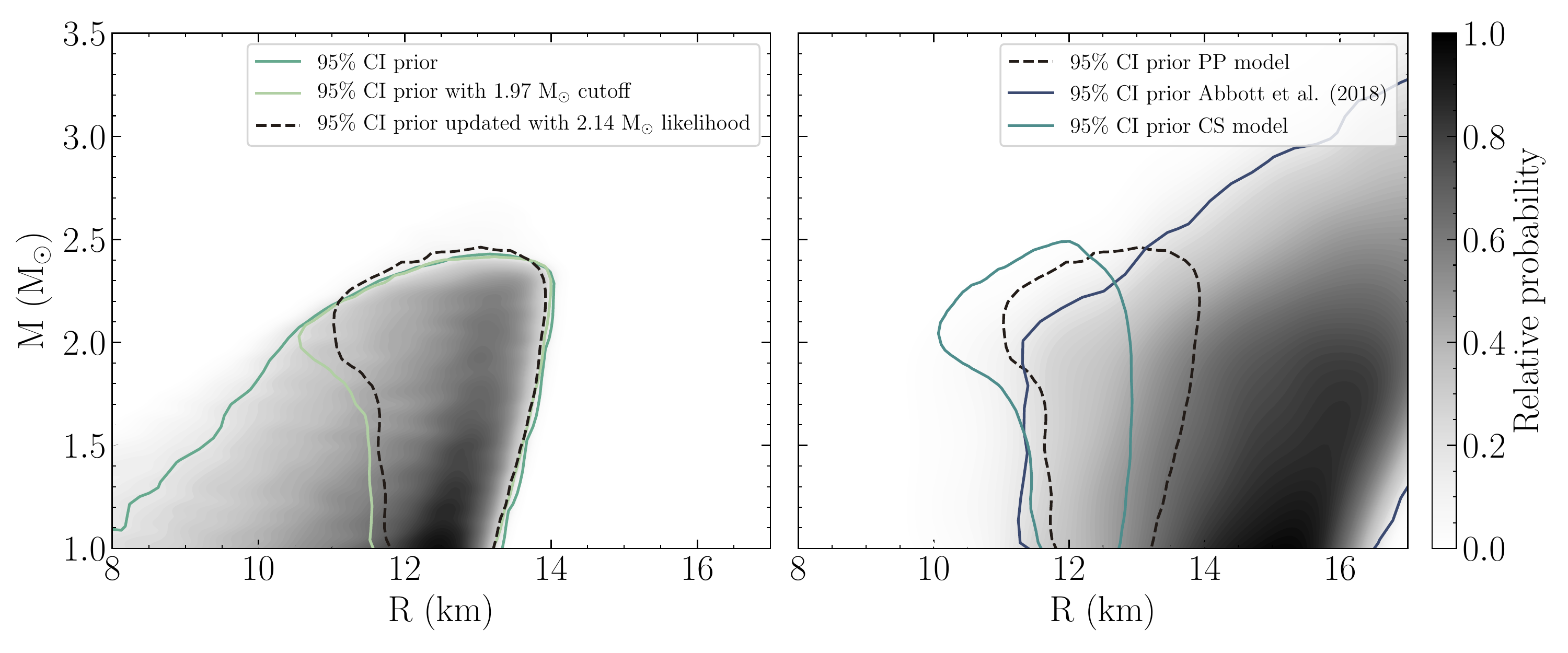}
\caption{Left panel: comparison between the full prior distribution (black shaded region, $95\%$ credible region bounded by the dark green contour) for the PP model with a $1.97$ M$_{\odot}$ cutoff (light green contour, as used in \citealt{Raaijmakers19}) and when updated by parameter estimation using the $2.14$ M$_{\odot}$ pulsar likelihood function (black dashed contour) from \citet{Cromartie19}. Using a cutoff in the prior allows for slightly smaller radii than using the likelihood function: this is due to both the higher mass of the center of the pulsar likelihood function and the fact that the likelihood function gives more weight to an EOS with a maximum mass of $2.14$ M$_{\odot}$ than, e.g., $2.05$ M$_{\odot}$.
Right panel: comparison between the $95\%$ credible regions of the prior distributions of the PP model (black, dashed contour) and the CS model (dark green contour), when updated with the $2.14$ M$_{\odot}$ pulsar likelihood function. We also show the $95\%$ credible region (black shaded region with blue contours) of the prior distribution using the spectral model \citep{Lindblom18} that was used in \citet{Abbott18}. From the comparison it is clear that the prior with the spectral model allows for much stiffer EOS but has a similar bound at low radii.}
\label{fig:fig2}
\end{figure*}

\section{Inference Framework}

We adopt the framework outlined previously in \citet{Raaijmakers19} and \citet{Greif19}, which we will briefly summarize here, including details of some adjustments made to incorporate information from the GW data of GW170817. Note that the Bayesian methodology is very similar to that outlined in \citet{Miller2019}, although there are differences in the prior assumptions (see Section~\ref{sec:disc}).

\subsection{Parameterizations} 
\label{subsec:param}

Two distinct parameterizations are considered: a three-piece polytropic (PP) model with varying transition densities between the polytropes \citep{Hebeler13}, and a speed of sound (CS) model based on physical considerations both at nuclear and high densities \citep{Greif19}. Both models are matched to an additional polytrope below $1.1 \, n_{0}$ (with saturation density $n_0=0.16 \, \text{fm}^{-3}$) with varying normalization that captures the range of allowed EOS calculated from chiral effective field theory (chiral EFT) interactions \citep{Hebeler10a, Hebeler13}. At densities below $0.5 \, n_0$ this polytrope is matched to the BPS crust EOS \citep{Baym71}. 

\subsection{Bayesian Parameter Estimation}

We use Bayesian methodology to estimate parameters in our EOS model. A more in-depth discussion on parameter estimation frameworks in the context of dense matter inference can be found in \citet{Riley18}, which we will very briefly describe here. Let us write, using Bayes' theorem, the posterior distributions of the EOS parameters $\bm{\theta}$ and central energy densities $\bm{\varepsilon}$ (together the \textit{interior} parameters) as 
\begin{equation}
\label{eq:eq1}
p(\bm{\theta}, \bm{\varepsilon} \,|\, \bm{d}, \mathcal{M})
\propto 
p(\bm{\theta} \,|\, \mathcal{M})
~
p(\bm{\varepsilon} \,|\, \bm{\theta}, \mathcal{M})
~
p(\bm{d} \,|\, \bm{\theta}, \mathcal{M}) \,,
\end{equation}
where $\mathcal{M}$ is the model that includes all assumed physics, and $\bm{d}$ is the data set from both \NICER observations and strain data of GW170817 from the GW detectors LIGO/Virgo. Note that the prior on the central energy densities $\bm{\varepsilon}$ is dependent on the EOS parameters $\bm{\theta}$, because the maximum stable central energy density is different for each set of $\bm{\theta}$. Given that the two observations are independent, we can separate the likelihood function as\footnote{Note that for notational simplicity, we omit conditional arguments that would denote the model used by a collaboration. The global model $\mathcal{M}$ can be considered as a proper superset of the union of these models.}
\begin{equation}
\label{eq:eq2}
\begin{split}
p(\bm{\theta}, \bm{\varepsilon} \,|\, \bm{d}, \mathcal{M})
\propto 
& ~ p(\bm{\theta} \,|\, \mathcal{M})
~
p(\bm{\varepsilon} \,|\, \bm{\theta}, \mathcal{M})\\
& ~~ \times p(\Lambda_1, \Lambda_2, M_1, M_2 \,|\, \bm{d}_{\rm GW})\\
& ~~ \times p(M_3, R_3 \,|\, \bm{d}_{\rm NICER})\
\end{split}
\end{equation}
where the nuisance-marginalized likelihood functions of (i) $M_3$ and $R_3$, and (ii) $\Lambda_1, \Lambda_2, M_1$, and $M_2$, are equated\footnote{For \NICER, the nuisance-marginalized likelihood function $p(\bm{d}_{\rm NICER} \,|\, M, R)\propto p(M, R \,|\, \bm{d}_{\rm NICER})$ because the joint prior $p(M, R)$ was flat \citep{Riley19}. For GW170817 the nuisance-marginalized likelihood function $p(\bm{d}_{\rm GW} \,|\, M_{1}, M_{2}, \Lambda_{1}, \Lambda_{2})\propto p(M_{1}, M_{2}, \Lambda_{1}, \Lambda_{2}, \,|\, \bm{d}_{\rm GW})$ because the priors used in \cite{GW170817sourceproperties} are flat in both masses and tidal deformabilities.} to the nuisance-marginalized joint posterior density distributions inferred by \citet{Riley19} and \citet{LIGO_161218_catalog}, respectively. The marginal GW likelihood function is degenerate under exchange of binary components, but we adopt the same convention as in \citet{GW170817sourceproperties} and define $M_1 \geq M_2$. The interior parameters $\bm{\theta}$ and $\bm{\varepsilon}$ (now containing three central energy densities, corresponding to one observed star by \NICER and two by the LVC) map deterministically to the parameters $M$, $R$, and $\Lambda$ (where we have assumed the rotation of the star, $\Omega$, to be zero\footnote{See \citet{Raaijmakers19} for a discussion on the spin of PSR J0030+0451. We assume that the spins of the two components in GW170817 are consistent with the spins of observed Galactic binary NS systems that will merge within a Hubble time. Therefore we use the posterior distribution of GW170817 in the case of a low-spin prior, i.e., $\chi <0.05$, which corresponds to slower rotation than PSR J0030+0451.}), allowing us to sample from the prior distribution of $\bm{\theta}$ and $\bm{\varepsilon}$ and then numerically evaluate the likelihood functions using kernel density estimation (KDE) on the posterior samples. We then draw from the joint posterior distribution $p(\bm{\theta}, \bm{\varepsilon} \,|\, \bm{d}, \mathcal{M})$ of all interior parameters.

However, one complication that arises when performing KDE on samples from the joint posterior distribution of the two masses associated with GW170817, is that, due to the extreme accuracy to which the chirp mass M$_{\rm{chirp}} = (M_1 M_2)^{3/5}/(M_1+M_2)^{1/5}$ is known relative to the uncertainty in the individual masses, the choice of bandwidth is difficult to make (for GW170817, $M_{\rm chirp}=1.186 \pm 0.001$~M$_{\odot}$; \citealt{GW170817sourceproperties}). A small bandwidth is necessary to accurately describe the chirp mass, while a larger bandwidth is necessary to smooth out finite sampling noise in the distribution of masses. Another complication is that when the two sampled central densities are uncorrelated---except for the assumption of a shared EOS---it is computationally expensive for samplers to find the region in the space of masses where all of the probability density is concentrated. 

To avoid these complications, and at the same time utilize the small chirp mass uncertainty, we fix it to its median value of $M_{\rm chirp}=1.186$~M$_{\odot}$. Consequently, the mass of the secondary object is a deterministic function of the mass of the primary object, and there is one fewer free central density parameter in the vector $\bm{\varepsilon}$. For details on the approximation invoked here, we refer to Appendix~\ref{appendix}, up to Equation~(\ref{eqn:delta-approximated GW likelihood}). For likelihood evaluation we use the mass ratio $q = M_2/M_1$ instead of the individual masses, transforming Equation (\ref{eq:eq2}) to \footnote{Note that, by transforming the parameters $M_1$ and $M_2$ to the mass ratio $q$ and $M_{\rm{chirp}}$, the LVC prior $p(q, M_{\rm{chirp}})$ is no longer flat. The likelihood function we approximated as proportional to the posterior density $p(\Lambda_1, \Lambda_2, q \,|\, \bm{d}_{\rm GW})$ is therefore contaminated. We have checked, however, that reweighting the GW170817 posteriors such that the prior $p(q,M_{\rm{chirp}})$ is flat has an unimportant effect on the likelihood function used for EOS inference. Moreover, note that approximating the conditional distribution $p(\Lambda_1, \Lambda_2, q \,|\, \bm{d}_{\rm GW}, M_{\rm{chirp}})$ by the marginal distribution $p(\Lambda_1, \Lambda_2, q \,|\, \bm{d}_{\rm GW})$ has a similarly unimportant effect.}

\begin{equation}
\label{eq:eq3}
\begin{split}
p(\bm{\theta}, \bm{\varepsilon} \,|\, \bm{d}, \mathcal{M})
\propto 
& ~ p(\bm{\theta} \,|\, \mathcal{M})
~
p(\bm{\varepsilon} \,|\, \bm{\theta},\mathcal{M})\\
& ~~\times p(\Lambda_1, \Lambda_2, q \,|\, \bm{d}_{\rm GW})\\
& ~~ \times p(M_3, R_3 \,|\, \bm{d}_{\rm NICER}).
\end{split}
\end{equation}
 Moreover, the deformability $\Lambda_{2}=\Lambda_{2}(\bm{\theta};q)$.
 
 In Figure~\ref{fig:fig1} \footnote{See the Zenodo repository \citet{plotdata} for the data and code to recreate all plots in this Letter.}we compare the updated prior distribution after including information from GW170817 with and without fixing the chirp mass. When the chirp mass is considered as a free parameter we include it as an element of $\bm{\theta}$,\footnote{Thus now mixing interior parameters and exterior spacetime parameters.} sample from a uniform prior $M_{\rm chirp}~\sim~U(1.180, 1.192)$, and define the likelihood function as $p(\Lambda_1, \Lambda_2, q, M_{\text{chirp}}\,|\, \bm{d}_{\rm GW})$, thus requiring four-dimensional KDE. The two distributions are, due to the small uncertainty in the chirp mass, almost equal, apart from some finite sampling noise. In the following, we fix the chirp mass to $M_{\rm chirp}=1.186$~M$_{\odot}$ in order to reduce the dimensionality of the parameter vector $\bm{\theta}$. 

We use the nested sampling software \MultiNest to draw weighted samples from the posterior distribution of $\bm{\theta}$ \citep{Feroz08, Feroz09, Feroz13, Buchner14}.

\begin{figure*}[t!]
\centering
\includegraphics[width=\textwidth]{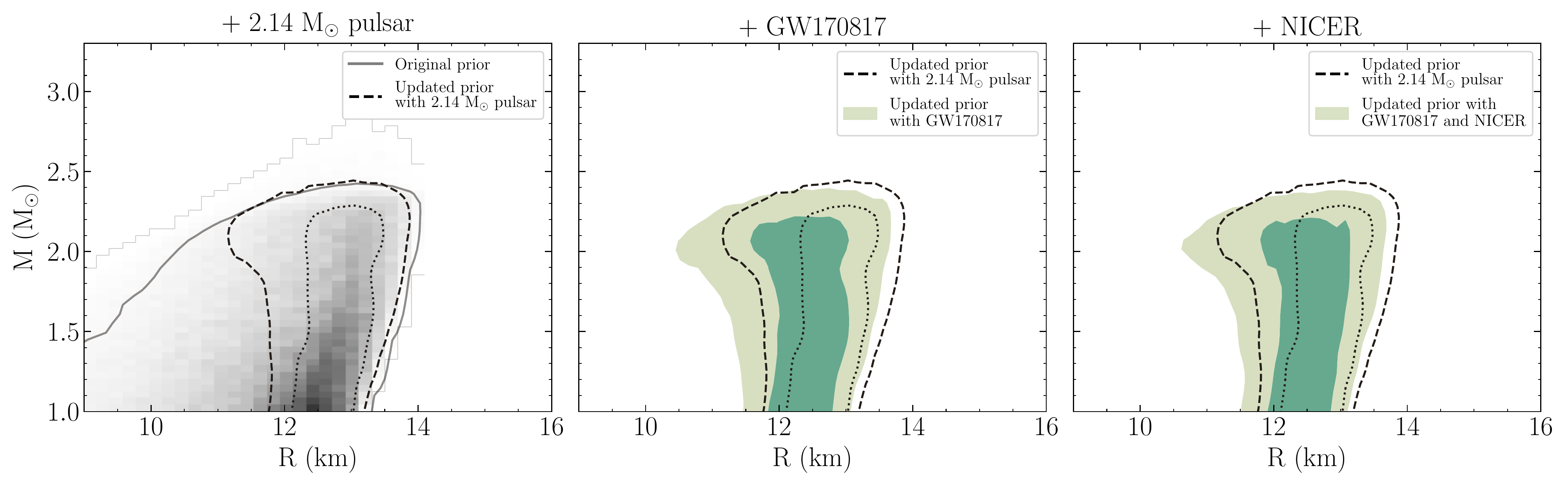}
\includegraphics[width=\textwidth]{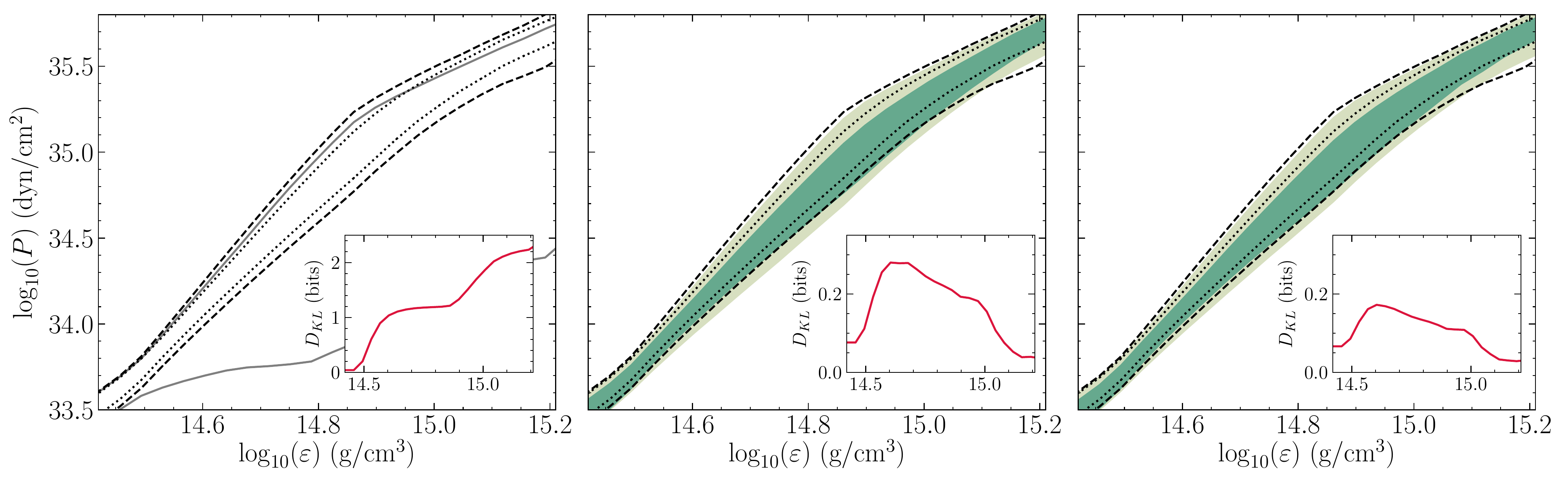}
\caption{Posterior distributions conditional on the PP model and given: (i) the $2.14$~M$_{\odot}$ pulsar alone (left panels), (ii) inclusion of the GW170817 measurements (middle panels), and (iii) inclusion of the mass and radius of PSR J0030+0451 inferred by \citet{Riley19} given \NICER data (right panels). In the top row we show how the posterior distributions update the prior distributions, by drawing a new central density given the inferred distribution on EOS parameters, $p(\varepsilon_c\,|\,\text{EOS})$. This is then transformed to the space of masses and radii, with the contours indicating the $68\%$ and $95\%$ credible intervals. In the bottom row we show the marginal posterior distributions of the pressure $P$ conditional on energy density $\varepsilon$, i.e., $p(P\,|\,\varepsilon, \bm{d}, \mathcal{M})$. The bands show the connected $68\%$ and $95\%$ credible intervals at each energy density $\varepsilon$. The grey lines in the left panels show the $95\%$ credible interval of the full prior, while the black dotted and dashed lines in all panels show the $68\%$ and $95\%$ credible regions of the updated prior when including information from the $2.14$~M$_{\odot}$ pulsar. The green contours show the same credible regions, but for posterior distributions that are inferred from multiple measurements of neutron star observables. In the lower right inset panels we illustrate the evolution of the Kullback-Leibler divergence as a function of energy density. We conclude that most information is gained from including the $2.14$ M$_{\odot}$ pulsar. The binary merger GW170817 favours softer EOSs than the prior, but the measured radius from PSR J0030+0451 favors stiffer EOSs, resulting in a final posterior distribution very similar to the prior.}
\label{fig:fig3}
\end{figure*}

\begin{figure*}[t!]
\centering
\includegraphics[width=\textwidth]{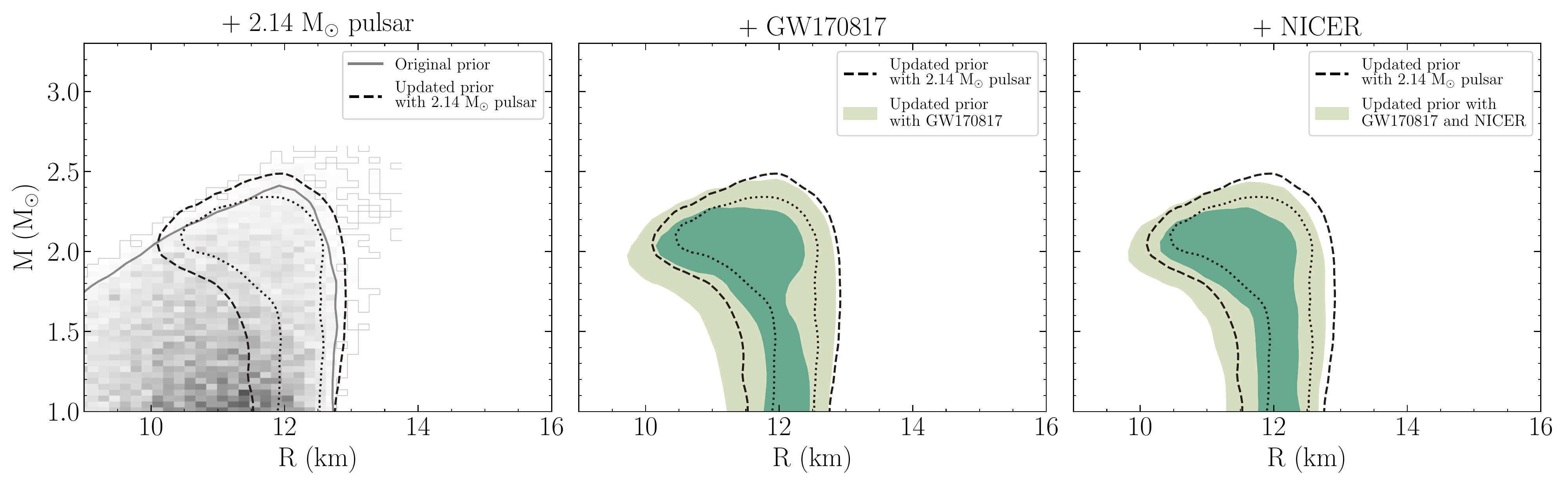}
\includegraphics[width=\textwidth]{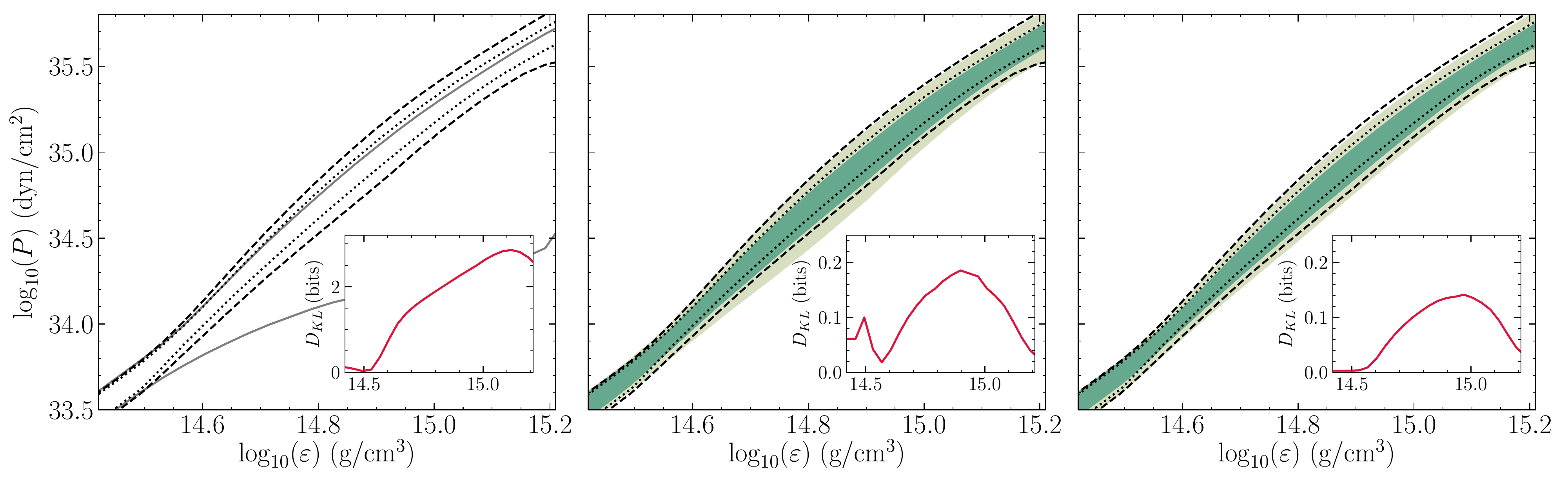}
\caption{Same as in Figure~\ref{fig:fig3} but for the CS model. Again we conclude that constraints from GW170817 point to softer EOSs with lower radii, but the results from \NICER point to stiffer EOS with higher radii. The final posterior distribution, conditional on the three different measurements combined, is then very similar to the distribution with only information from the 2.14~M$_{\odot}$ pulsar included.}
\label{fig:fig4}
\end{figure*}

\subsection{Priors}

The bounds of the prior ranges of parameters used in this analysis are identical to those discussed in Section~2.1.1 of \citet{Raaijmakers19} and Section~3.1.1 of \citet{Greif19}. Within these bounds all EOS parameters are sampled uniformly, while the central density of a star, $\varepsilon_c$, is uniformly sampled in logarithmic space between $\log({\varepsilon_c}) = 14.6$ and an upper bound that is determined by the maximum central density of each particular EOS. We consider an EOS in the PP model up until the highest density that corresponds to a stable NS or to the point where causality is no longer satisfied, i.e., where the speed of sound exceeds the speed of light, $c_s>c$. The CS model has slightly more restrictive requirements: 
\begin{enumerate}[(i)]
\item The speed of sound for all densities should be lower than the speed of light. 
\item At asymptotic densities ($\sim\!50 n_0$) the speed of sound should converge to $c_s = \sqrt{1/3} c$ from below, based on theoretical calculations of $c_s$ in the framework of perturbative quantum chromodynamics \citep{Fraga14}.
\item At low densities the speed of sound can be described by that of a normal Fermi liquid such that we require $c_s \leq \sqrt{0.163} \, c$ at densities below $1.5 \, n_0$ \citep[for more details, see][]{Greif19}.
\end{enumerate}

A notable change from the prior used in \citet{Raaijmakers19} is how we implement information from pulsar mass measurements in our analysis.\footnote{We refer the reader to the discussion in section 4.2 of \citet{Raaijmakers19}, and to the arguments in section 4.1 of \citet{Miller2019}.} These high-precision measurements obtained from the timing of radio pulsars restrict softer EOSs by requiring each EOS to be able to support the heaviest NSs. There have been several massive NS detected, with the most stringent constraints coming from PSR J0348$+$0432 with a mass of $2.01_{-0.04}^{+0.04}$~\msol \citep{Antoniadis13} and more recently PSR J0740$+$6620 with a mass of $2.14_{-0.09}^{+0.10}$~\msol \citep{Cromartie19}. In many previous analyses the lower $1\sigma$ limit of such a mass measurement was taken as the minimum mass that an EOS has to support \textit{a priori}. However, \citet{Miller2019,Miller19} do not make such an assumption and emphasize that likelihood information about high-mass pulsars be treated accurately \citep[see also][]{Alvarez16}. Here we approximate the highest pulsar mass measurement as a Gaussian likelihood function,\footnote{A value of $0.09$~M$_{\odot}$ was chosen for $\sigma$ in the Gaussian likelihood function, which is not representative for the upper tail of the pulsar mass distribution. We believe, however, that the effect on the posterior distributions of the EOS parameters is small enough to justify this approximation.} such that Equation (\ref{eq:eq3}) reads
\begin{equation}
\label{eq:eq4}
\begin{split}
p(\bm{\theta}, \bm{\varepsilon} \,|\, \bm{d}, \mathcal{M})
\propto 
& ~ p(\bm{\theta} \,|\, \mathcal{M})
~
p(\bm{\varepsilon} \,|\, \bm{\theta}, \mathcal{M})\\
&~~\times
p(\Lambda_1, \Lambda_2, q \,|\, \bm{d}_{\rm GW})\\
&
~~\times p(M_3, R_3 \,|\, \bm{d}_{\rm NICER})\\
&
~~\times p(M_4 \,|\, \bm{d}_{\rm radio}).\\
\end{split}
\end{equation}
The vector $\bm{\varepsilon}$ now contains a fourth central energy density (drawn from the same prior as described at the beginning of this section) for which the corresponding mass, $M_4$ is used to evaluate the Gaussian likelihood function.
We compare the effect of a cutoff in the prior with the implementation of the new likelihood function for the PP model in the left panel of Figure~\ref{fig:fig2}. The solid lines indicating the $95\%$ credible region show that the prior distribution between the two methods is very similar, although the likelihood implementation of the $2.14$~\msol pulsar is slightly more constraining at lower radii due to the higher mass. In the right panel of Figure~\ref{fig:fig2} we compare the prior distributions for the two parameterizations used in this Letter with the prior distribution of the spectral parameterization used in \citet{Abbott18}. The spectral parameterization allows for much larger radii than we consider here, due to using only a crust EOS without implementing nuclear physics constraints around nuclear saturation density. This is taken into account in this work \citep[see also][]{Hebeler13} by adopting the EOS band based on chiral EFT up to $1.1 n_0$. The exact breakdown density of chiral EFT is not fully known, but many calculated and also predicted nuclear properties are consistent with experiment \citep{Hebeler15}, including the symmetry energy and other matter properties at saturation density, suggesting that the range of possible EOSs predicted by chiral EFT is valid up to around nuclear saturation density \citep[see also Section~4.2 of][]{Raaijmakers19}. 

\subsection{Generalization to a Large Number of Stars}

The separation of the likelihood function based on different observables in Equation (\ref{eq:eq4}) is a useful way of analysing multiple sources at the same time. In the future, however, when a population of neutron stars is observable, this can quickly become computationally intractable---depending on the sampling algorithm applied---as the number of densities constituting the parameter vector $\bm{\varepsilon}$ grows linearly with the number of observed stars. One can perform the parameter estimation sequentially in this case, where the prior $p(\bm{\theta} \,|\, \mathcal{M})$ is updated after each iteration and sampled from in the next (see also Figure~2.1 in \citealt{Riley18}).\footnote{Although natural, sequential updating does not come without its own technical challenges. Moreover, in a rigorous population-level context, one should consider the role of hyperparameters.}

\section{EOS Constraints Given Multi-messenger Observations}

For both parameterizations discussed in Section~\ref{subsec:param} we draw weighted samples from the posterior distribution $p(\bm{\theta}~|~\bm{d}, \mathcal{M})$ using nested sampling. In order to explore the effect of different measurements on this posterior distribution, we start by only considering information from the 2.14~M$_{\odot}$ pulsar; we then include information from the binary merger GW170817, and finally we include the more recent \NICER measurements of PSR J0030+0451. The GW data we use here are the publicly available posterior samples~\cite{LIGODATA} which assume certain priors on the GW parameters as described in~\citet{LIGO_161218_catalog}; a study of the impact of changing these priors is outside the scope of this work but see~\citet{Lattimerinprep}.

\subsection{EOS Constraints Assuming GW170817 Was a NS-NS}

We illustrate the posterior distribution in two different ways in Figures~\ref{fig:fig3} and \ref{fig:fig4}. The lower panels show the $68\%$ and $95\%$ credible intervals on the pressure at each energy density given the inferred distribution on EOS parameters, i.e., $p(P\,|\,\varepsilon, \bm{d}, \mathcal{M})$. The upper panels show the updated prior distribution for a new star given the inferred distribution on EOS parameters, transformed to the space of mass and radii. More practically, this means that for each EOS in the posterior distribution $p(\bm{\theta} \,|\, \bm{d}, \mathcal{M})$ we draw a new central density from the prior $p(\bm{\varepsilon} \,|\,\bm{\theta})$, where we use again a uniform prior in log space with the upper bound determined by the maximum stable neutron star of that EOS. From this central energy density we then compute the corresponding mass and radius pair. By combining all pairs for each EOS sample in the posterior we obtain a distribution of masses and radii. The contours again indicate the $68\%$ and $95\%$ credible regions. Visually inspecting the posterior distributions for both the PP model (Figure~\ref{fig:fig3}) and the CS model (Figure~\ref{fig:fig4}) indicates that the inferred masses and tidal deformabilities for GW170817 favor softer EOSs than our prior. Folding in information about the radius of PSR J0030+0451, however, with a peak value around $12.7$~km, favors stiffer EOS. As a result the final posterior distribution is only slightly shifted toward smaller radii but otherwise closely follows the distribution when only the highest-mass pulsar is included. We also note that one should be careful when inferring the maximum mass allowed for a NS from Figures \ref{fig:fig3} and \ref{fig:fig4} because the high-mass end of these distributions depends sensitively on the prior on central energy densities. See also Section \ref{sec:maxmasses} and Figure \ref{fig:fig6} where we explicitly show the distribution of the maximum mass by taking the highest allowed central energy density from each EOS instead of sampling from the prior on $\varepsilon$.

In order to quantify this we compute the Kullback-Leibler (KL) divergence \citep{kullback1951} between the two distributions shown in the lower panels of Figure~\ref{fig:fig3} and \ref{fig:fig4} at a given energy density $\varepsilon$ (see lower right inset panels). The KL divergence is an asymmetric measure of how one probability distribution differs from another; when computed using a logarithm of base two, the divergence has units of bits. As expected, most of the information is gained from folding in the 2.14~M$_{\odot}$ pulsar constraint. The posterior distribution given GW170817 alone exhibits greater divergence from the prior than does the posterior distribution given GW170817 \textit{and} \NICER information; that said, both divergences are small at all densities.  

\begin{table}[t!]
\label{tab:table1}
\begin{tabular}{@{}llll@{}}
\toprule
 Posterior  & \multicolumn{2}{c}{log(Z)}        & K  \\ \midrule
                        & PP model        & CS model        &               \\
\midrule
+ 2.14 M$_{\odot}$ pulsar & $-1.69$ $\pm$ 0.03  & $-2.22$ $\pm$ 0.02  &   0.70 \ \ \ \  \\
+ GW170817              & $-15.44$ $\pm$ 0.02 & $-15.08$ $\pm$ 0.02 &   1.70      \\
+ \NICER                 & $-17.05$ $\pm$ 0.03 & $-17.10$ $\pm$ 0.03 & 1.05   \\ \bottomrule
\end{tabular}
\caption{Log-evidences (Z) for the three posterior distributions and two parameterizations. Also quoted are the Bayes' factors (K), computed as the ratio of the evidence for the PP model over the evidence for the CS model. Following the interpretation of \citet{KassRaft95} there is no significant support for one parameterization over the other. }
\end{table}

Finally, we compute the Bayes' factors to investigate whether one parameterization is favored over the other by the data. Assuming the two discrete models to have equal probability \textit{a priori}, the Bayes' factor reduces to the ratio of the evidences of the two posteriors. We quote the values for the three different posterior distributions in Table (\ref{tab:table1}), where the Bayes' factor $K$ is the ratio of the PP model over the CS model. To interpret the values of $K$ we follow the table of \citet{KassRaft95} and conclude that none of the Bayes' factors shows substantial support for one of the models over the other. 

\begin{figure*}[t!]
\centering
\includegraphics[width=.48\textwidth]{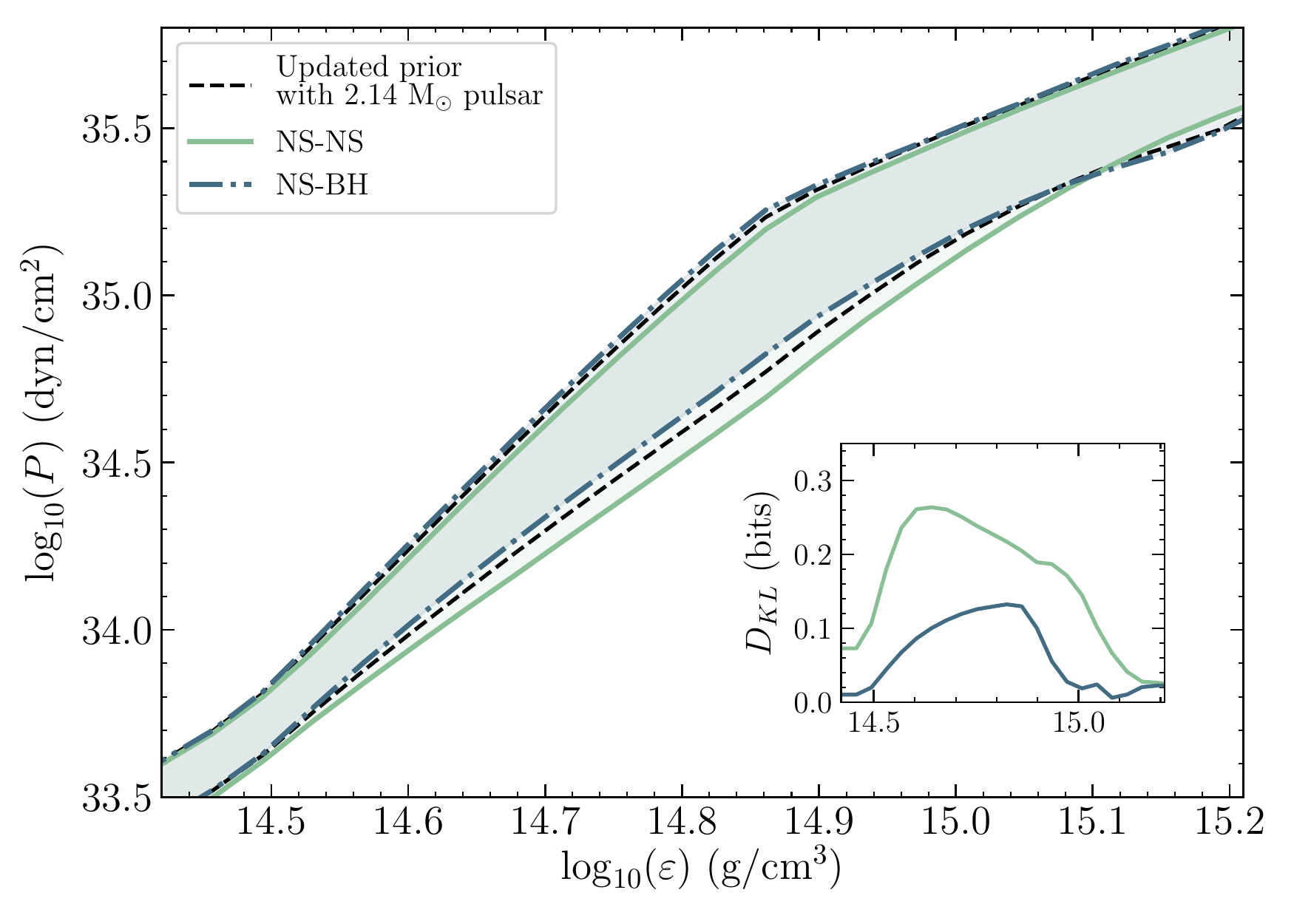}
\includegraphics[width=.48\textwidth]{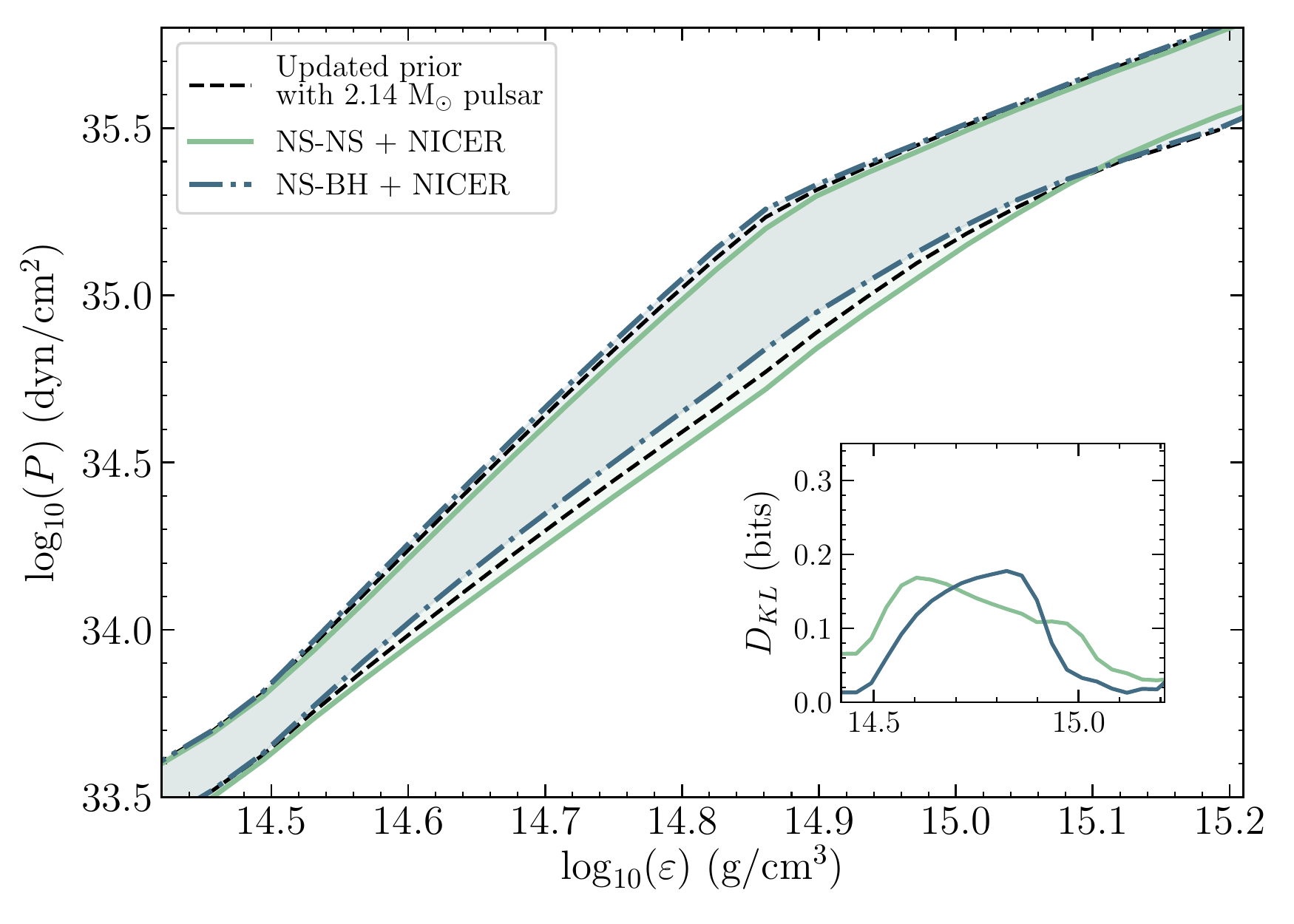}
\caption{Comparison between the posterior distributions obtained for the assumption that GW170817 is a NS-NS or a NS-BH merger, without (left panel) and with (right panel) the \NICER measurements of PSR J0030+0451. Here we only show results for the PP model. The lines show the connected $95\%$ credible regions at a given energy density $\varepsilon$. The black lines show the same credible regions when the posterior distribution is only informed by the 2.14~M$_{\odot}$ pulsar.  The lower right inset illustrates the evolution of the KL divergence with energy density $\varepsilon$ and indicates that GW170817 is slightly more constraining for the EOS when assumed to be a NS-NS merger.}
\label{fig:fig5}
\end{figure*}

\subsection{EOS Constraints Assuming GW170817 Was a NS-BH}
\label{sec:BHNS}
Based on the gravitational wave signal and observed electromagnetic counterpart from GW170817 there is a non-negligible chance that one of the compact objects involved in the merger was a light BH \citep{Yang:2017gfb,Ascenzi19, Coughlin19,Hinderer19}, provided that the objects had an unequal mass ratio. Such a light BH could for example be formed during an earlier merger of two NSs, or originate from density fluctuations in the early universe \citep{Garcia96}. 

We show that making different assumptions related to the nature of the merger can affect the inferred EOS. More specifically we investigate the impact of GW170817 being a NS-BH by fixing the tidal deformability of the heavier object \textbf{$\Lambda_1$} to zero. One complication, however, is that the posterior samples provided in \citet{GW170817sourceproperties} do not contain enough samples when we restrict $\Lambda_1 = 0$. Instead, we perform a coordinate transformation on the posterior samples to the effective tidal deformability $\Tilde{\Lambda}$, a combination of the two independent tidal deformabilities and masses. To incorporate the constraints on the NS-BH scenario from electromagnetic (EM) observations of the associated kilonova we follow the approach outlined in \citet{Hinderer19}. We use the model of \citet{Foucart18} to compute the remnant mass outside the BH after the merger $M_{\rm rem}$ from the progenitor NS compactness, which is related to its tidal deformability $\Lambda_2$, the dimensionless spin of the BH $\chi_1$, and the mass ratio. For the spin we only consider aligned or anti-aligned spins with the orbital angular momentum. We obtain these inputs from the GW data by transforming the posterior samples of GW170817 to $\Tilde{\Lambda}$ and $\chi_{\rm{eff}}$, the effective dimensionless spin, setting the BH tidal deformability $\Lambda_1 = 0$ and the NS spin $\chi_{2}=0$. We apply a conservative cut to the GW170817 posterior by only considering samples that have $M_{\rm{rem}}>0.1$ \msol, which can produce the observed bolometric light curve of GW170817 \citep{Kasliwal17} within some error margins based on semi-analytic light curve modeling. Effectively this is a crude approximation to the likelihood function that is zero if $M_{\rm{rem}}$ is below this threshold and uniform if above. The posterior distribution of the EOS parameters is then given by

\begin{equation}
\label{eq:eq5}
\begin{split}
p(\bm{\theta}, \bm{\varepsilon} \,|\, \bm{d}, \mathcal{M})
\propto
& ~ p(\bm{\theta} \,|\, \mathcal{M})
~
p(\bm{\varepsilon} \,|\, \bm{\theta}, \mathcal{M})\\
& ~~
\times p(\Tilde{\Lambda}(\Lambda_2, M_1, M_2), q \,|\, \bm{d}_{\rm GW,EM})\\
& ~~ \times p(M_3, R_3 \,|\, \bm{d}_{\rm NICER})\\
& ~~
\times p(M_4 \,|\, \bm{d}_{\rm radio}) \,,
\end{split}
\end{equation}
in which the tidal deformability of the heavier object is fixed to zero and the chirp mass is again fixed to $M_{\rm chirp}=1.186$~M$_{\odot}$. Equation~(\ref{eq:eq5}) requires that the joint prior density of $\Tilde{\Lambda}$ and $q$ is \textit{flat}. However, the prior density is \textit{not} flat. In particular, the prior distribution of  $\Tilde{\Lambda}$ exhibits undesirable behavior near zero, where prior support drops to zero. To minimize the induced bias we reweight the posterior samples to a flat prior in $\Tilde{\Lambda}$ using the method described in \citet{GW190425}. The difference is, however, small since the EM information already excludes values of $\Tilde{\Lambda}$ close to zero. Finally we note that the term $p(\Tilde{\Lambda}(\Lambda_2, M_1, M_2), q \,|\, \bm{d}_{\rm GW,EM})$ is marginalized over $\Lambda_1$ and therefore the BH-NS likelihood function is contaminated with the likelihood of a system with two deformable objects. However, comparing the marginalized distribution with the conditional distribution  $p(\Tilde{\Lambda}(\Lambda_2, M_1, M_2), q \,|\, \bm{d}_{\rm GW,EM}, \Lambda_1)$ shows that the approximation is valid here in the context of illustrating the impact of different assumptions about the binary (see Appendix (\ref{AppendixB})).

In Figure~\ref{fig:fig5} we compare the posterior distribution for the EOS for the assumption that GW170817 is a NS-NS or NS-BH merger, with the prior distribution obtained from only the $2.14$~M$_{\odot}$ pulsar measurement. In the left panel we only consider GW170817 and the pulsar mass, while in the right panel we also include the \NICER measurement. Based on the KL divergence as a function of energy density, we conclude that in both cases GW170817 has more support for softer EOSs when assumed to be a double NS merger. This can be explained by the fact that the relatively low inferred value of $\Tilde{\Lambda}$ in \citet{GW170817sourceproperties} can be achieved with a higher value of $\Lambda_2$ (i.e., stiffer EOS) when $\Lambda_1 = 0$ and the fact that the NS-BH scenario is only consistent with the EM counterpart for unequal mass ratios and larger NSs.

\section{Discussion}
\label{sec:disc}

In this work we have analyzed the combined constraints on the dense matter EOS given the recent inferred mass and radius of PSR J0030+0451 by \citet{Riley19} using \NICER data, and the measurement of the gravitational wave signal from GW170817, in combination with the radio measurement of a $2.14$~M$_{\odot}$ pulsar.

\subsection{Multi-messenger Contributions}

The posterior distributions show that the most information is gained from the most massive pulsar mass measurement. In combination with the restricted range of possible EOS at lower densities described by the chiral EFT band and, for the CS model, by the approximation of NS matter as a Fermi liquid, the pulsar mass already puts stringent constraints on the EOS. When including information from GW170817, softer EOSs are yielded \textit{a posteriori}, but this is only a small effect because EOSs consistent with the $2.14$~M$_{\odot}$ pulsar mass measurement are restricted by causality to higher radii. Finally, the recent \NICER measurement shows more support for stiffer EOS, causing the posterior distribution to have a narrow peak where the likelihood functions of \NICER and GW170817 overlap with the information from the $2.14$~M$_{\odot}$ pulsar and the chiral EFT band.

In order to quantitatively assess the prior-to-posterior information gain through sequential multi-messenger updates, we have computed KL divergences as a function of energy density. The divergences indicate that most information is gained from the radio pulsar mass measurement. The information gain from GW170817, given prior radio information, is small. Further, including the \NICER likelihood yields a smaller information gain because the radio and X-ray modeling constrain a similar part of the EOS parameter space, i.e., provide less support for softer EOSs.

Note that these divergences all depend on which distribution is compared with which. For example, comparing the constraints from \NICER or GW170817 with the original, more diffuse prior distribution would yield a higher KL divergence. We argue, however, that a logical, unique order of precedence would be ideal. An obvious option is to chronologically incorporate the various (astronomical) measurements that constrain the EOS.

It is however difficult to design an update order that is truly chronological, given that (i) not all constraints are compiled in any one analysis, and (ii) multi-messenger constraints are being derived contemporaneously, both given newly acquired data, and given archival data when the modeling procedure is revolutionized. Typically there will not be a clear chronology, and even if there were, we might try to account for the number of physical assumptions a constraint is conditional on. If we assume that the number of assumptions anti-correlates with robustness to systematic error in our models of reality, we can attempt to compile information---and archive constraints---in loose order of robustness.\footnote{In principle, if a chronologically earlier constraint is biased, future unbiased, informative constraints should offer a strong opinion. Thus bias should resolve in time, provided that computation can be performed accurately \citep{Raaijmakers19}. This is especially true if models are revolutionized and used to reanalyze data, at the expense of a clear chronology. In practice, however, it is not always straightforward to accurately update knowledge with future information when resolution would be required in the tail of a prior distribution.}

We therefore opt to start with information contributed from radio timing of pulsars in relativistic binaries. There are multiple constraints to consider: constraints for two systems were reported before GWs were first detected \citep{Demorest10,Antoniadis13}, with measurements for the very first being updated in recent years with continued timing \citep{Fonseca16, Arzoumanian18, Cromartie19, Miller2019}. It is believed that radio measurements relying solely on the relativistic Shapiro delay are robust to systematic error. However, the first report of a pulsar with a mass above 2~M$_{\odot}$ is dependent on theoretical models of white dwarf evolution \citep{Antoniadis13}. The issue of choice can of course be straightforwardly nullified by incorporating all of the radio pulsar mass measurements, each of which encodes orthogonal information in a population-level context. In this work we chose to use a single astronomical source for each class of astronomical messenger.
Of the two highly informative measurements relying solely on the relativistic Shapiro delay, derived by \citet{Arzoumanian18} and \citet{Cromartie19}, we chose to use the constraint reported by the latter.

\begin{figure*}[t!]
\centering
\includegraphics[width=.98\textwidth]{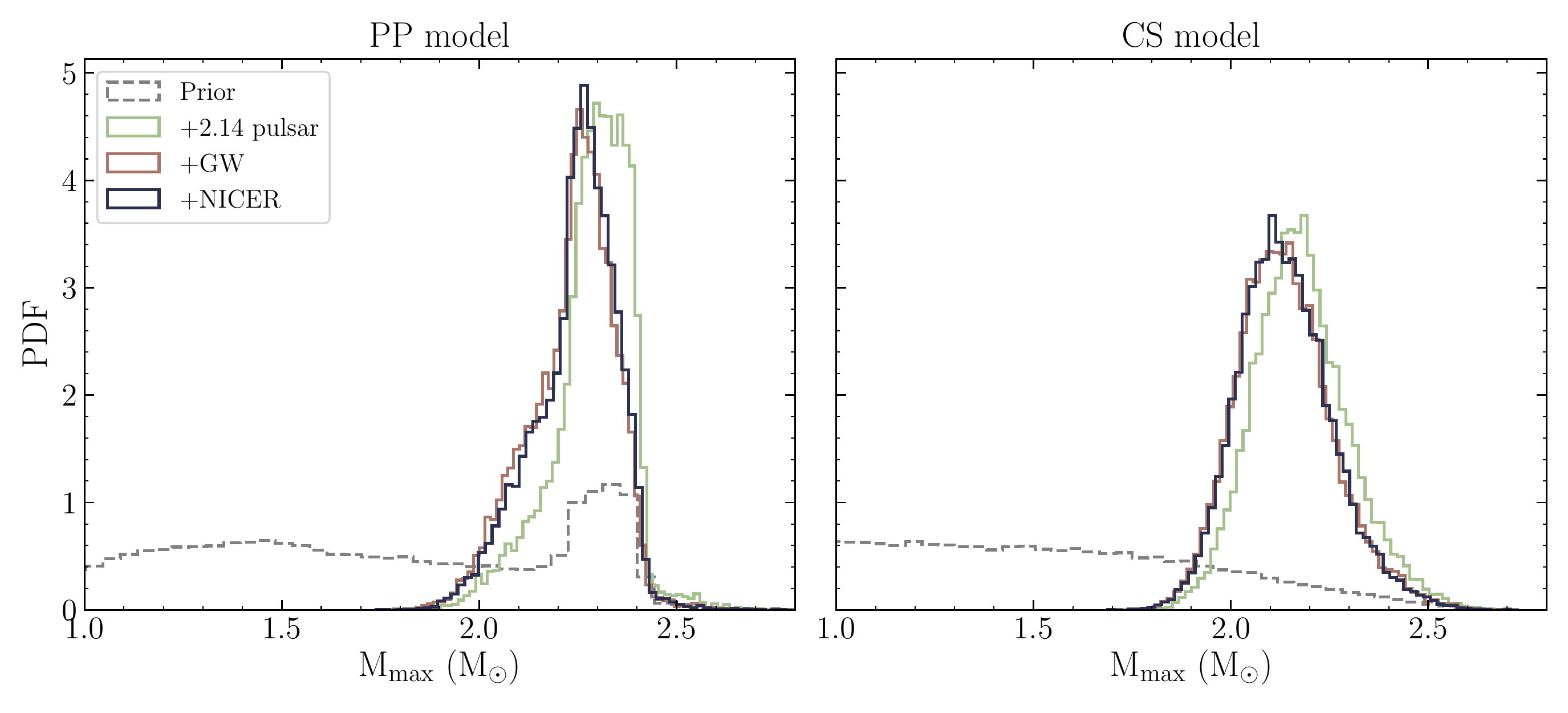}
\caption{Posterior distributions of the maximum mass of a non-spinning NS M$_{\rm{max}}$ that is supported by the inferred EOS after including the pulsar mass measurement, GW170817 and the results from \NICER. For the PP model we find a median value of M$_{\rm{max}} = 2.26^{+0.16}_{-0.24}$ \msol while for the CS model we find M$_{\rm{max}} = 2.13^{+0.26}_{-0.22}$ \msol, where the upper and lower limit bound the 95\% credible region.}
\label{fig:fig6}
\end{figure*}

\subsection{Implications for the Maximum Mass of NSs}
\label{sec:maxmasses}

The maximum mass of NSs places important constraints on the EOS and can aid in determining the nature of a compact binary merger and the following merger remnant. In Fig.~\ref{fig:fig6} we show the posterior distributions for M$_{\rm{max}}$ for the PP and CS parameterizations considered in this Letter. These distributions are obtained by calculating for each EOS sample in the posterior distribution $p(\bm{\theta} \,|\, \bm{d}, \mathcal{M})$ the maximum mass for that EOS. The gray dashed lines indicate the prior support on M$_{\rm{max}}$, which drops down rapidly above $\sim2.5$ \msol for both parameterizations. The bump around $2.3$ \msol for the PP model is a consequence of EOS that terminate at lower maximum masses when their sound speed reaches the speed of light before reaching a maximally stable NS. This leads to a higher value for the PP model of M$_{\rm{max}} = 2.26^{+0.16}_{-0.24}$ \msol compared to the CS model, M$_{\rm{max}} = 2.13^{+0.26}_{-0.22}$ \msol, when including information from pulsar mass measurements, GW170817, and \NICER. Here we quote median values and upper and lower limits that bound the $95\%$ credible region. We find that the maximum mass distribution presented here is broadly consistent with values found in other works 
\citep[see, e.g.,][]{Margalit17, Shibata17, Alsing18, Rezzolla18, Ruiz18, Tews19}. We find, however, a lower upper-limit since some of these analyses also include information from the kilonova light curves that suggest the formation of a metastable NS and disfavor the prompt formation of a BH.

\subsection{Comparison to Other Work}

We first compare our results with the analysis of \citet{Miller19}, who use a similar Bayesian approach to combine results from \NICER, GW170817, and radio pulsar measurements to constrain the EOS.
The parameterization of the EOS used in this work differs, however, in several aspects from the two parameterizations used by \cite{Miller19}. For the latter the crust EOS of \cite{Douchin01} is considered up to $0.5 \, \rho_s$. Beyond $0.5 \, \rho_s$, two different extrapolations to higher densities are used. One approach used by \cite{Miller19} is to implement the spectral parameterization by \cite{Lindblom10,Lindblom18}. The second parameterization is a piecewise polytropic expansion with two more polytropic segments than used in this work, totalling up to five segments. The range of the first polytropic index of \citet{Miller19} is chosen to be rather restrictive in the context of chiral EFT \citep{Hebeler13}. Moreover, \citet{Miller19} note that first-order phase transitions are not allowed in the case of the spectral parameterization, but they are permitted non-exhaustively in the case of the piecewise polytropic expansion (as in the present work). Overall the priors on their parameterizations, in particular the spectral model (see also Figure~\ref{fig:fig1}) allow for a larger range of possible EOS functions. Visually comparing our inferences with Figure~14 in \citet{Miller19}, however, we deem the posterior distributions to be consistent. 

Next, we compare to previous analyses of EOS constraints conditional on GW170817. The results of the LVC ~\citep{Abbott18} are broadly consistent with our analysis here, yet there are noticeable differences in the lower bound of the $90\%$ confidence interval in radius (comparing Figures~\ref{fig:fig3} and \ref{fig:fig4} with Figure~3 in \citealt{Abbott18}). We attribute this discrepancy primarily to the different assumptions on the speed of sound: the spectral EOSs in \citet{Abbott18} allowed models with up to $10\%$ violation of causality, while both parameterized EOSs used in our analysis were strictly causal with $c_s\leq c$. Additional differences are that \citet{Abbott18} use as the crust EOS a SLy model up to $\sim 0.5, \rho_{s}$ while we use the BPS model, and we incorporate nuclear physics constraints based on chiral EFT up to around saturation density \citep{Hebeler13}. The latter affects mainly the prior at the large radius end.

A number of independent analyses of the GW data also found broadly consistent results. \citet{De18} analyzed the GW data with the source location and distance fixed to those determined from the electromagnetic observations, identified scaling relations between tidal parameters and mass ratio using piecewise polytropic models, and determined EOS and radius constraints for a $1.4$~M$_{\odot}$ star to $8.9<R_{1.4}<13.2$~km consistent with our results here.  \citet{Most:2018hfd} computed a large catalog of piecewise polytropic EOSs, parameterizing both hadronic models and those with phase transitions. They analyzed the subset of these consistent with high-mass NSs and constraints on tidal deformability parameter $\tilde \Lambda$, both of which were imposed as hard cutoffs. Their results for the case $M_{\rm max}>2.01$~M$_\odot$ and $\tilde \Lambda_{1.4}<800$ (e.g., their Figure~1, top left panel) are consistent with our findings here. \citet{Capano:2019eae} used a different speed of sound parameterization based on similar chiral EFT constraints, as well as different priors, in particular a uniform distribution in radius. Thus, their results using the GW data alone are skewed more toward smaller radii $9.2 \lesssim R_{1.4}\lesssim 12.3$~km when imposing the chiral EFT limits up to $\rho_s$. \citet{Essick:2019ldf} used a nonparametric EOS inference under different priors, also finding broadly consistent results. 

In conclusion, the large statistical uncertainties in the available \NICER and LIGO/Virgo likelihood functions lead to broad agreements on the EOS and radius constraints across different analyses. However, the impact of priors, assumptions, and parameterizations is starting to become discernible, as we have shown. Highly anticipated upcoming observations with LIGO/Virgo, \NICER, and radio pulsars will constrain the EOS for populations of NSs, and yield unique insights into the properties of cold, dense matter. Our method can readily ingest the additional information from multiple sources, as well as incorporate new constraints from subatomic experiments and theory.

\acknowledgments
We thank the anonymous referee for comments that helped to improve this work. This work was supported in part by NASA through the \NICER mission and the Astrophysics Explorers Program. G.R., T.H., and S.N. are grateful for support from the Nederlandse Organisatie voor Wetenschappelijk Onderzoek (NWO) through the VIDI and Projectruimte grants (PI Nissanke). T.E.R. and A.L.W. acknowledge support from ERC Starting Grant No.~639217 CSINEUTRONSTAR (PI Watts). This work was sponsored by NWO Exact and Natural Sciences for the use of supercomputer facilities, and was carried out on the Dutch national e-infrastructure with the support of SURF Cooperative. S.K.G., K.H., and A.S. acknowledge support from the Deutsche Forschungsgemeinschaft (DFG, German Research Foundation) -- Project-ID 279384907 -- SFB 1245. S.G. acknowledges the support of the CNES. R.M.L. acknowledges the support of NASA through Hubble Fellowship Program grant HST-HF2-51440.001. J.M.L. acknowledges support by NASA through the NICER mission with Grant 80NSSC17K0554 and by the U.S. DOE through Grant DE-FG02-87ER40317. This research has made extensive use of NASA's Astrophysics Data System Bibliographic Services (ADS) and the arXiv.  We thank Jocelyn Read, Wynn Ho, and Cole Miller for comments on a draft manuscript.

\software{Python/C~language~\citep{python2007}, GNU~Scientific~Library~\citep[GSL;][]{Gough:2009}, NumPy~\citep{Numpy2011}, Cython~\citep{cython2011}, SciPy~\citep{Scipy}, MPI~\citep{MPI}, \project{MPI for Python}~\citep{mpi4py}, Matplotlib~\citep{Hunter:2007,matplotlibv2}, IPython~\citep{IPython2007}, Jupyter~\citep{Kluyver:2016aa}, \MultiNest~\citep{Feroz09}, \textsc{PyMultiNest}~\citep{Buchner14}.}

\bibliography{References}

\appendix
\section{GW Likelihood and Prior Implementation}\label{appendix}
\noindent 

\indent The posterior distribution derived by the LVC, assuming both binary components are NSs, is
\begin{equation}
p\left(\boldsymbol{M},\boldsymbol{\Lambda} \,|\, \boldsymbol{d}\right)
\propto
p\left(\boldsymbol{d} \,|\, \boldsymbol{M},\boldsymbol{\Lambda}\right)
\end{equation}
where $\boldsymbol{M}=[M_{1},M_{2}]^{\top}$,  $\boldsymbol{\Lambda}=[\Lambda_{1},\Lambda_{2}]^{\top}$, $\boldsymbol{d}$ is the strain data vector, and we omit the conditional argument representing the model. The joint prior density $p(\boldsymbol{M}, \boldsymbol{\Lambda})$ is flat on compact support, such that the posterior density is proportional to the nuisance-marginalized likelihood function $p\left(\boldsymbol{d} \,|\, \boldsymbol{M},\boldsymbol{\Lambda}\right)$. This function is symmetric under exchange of the binary components---i.e., $M_{1}\leftrightarrow{}M_{2}$ \textit{and} $\Lambda_{1}\leftrightarrow{}\Lambda_{2}$. We are therefore free to define one mass as being associated with the most massive component: $M_{i}\geq M_{j}$ for $i\neq j$. If we opt \textit{not} to, a numeric label is always associated with the same physical object. We are free to transform spaces, such that the likelihood function is $p\left(\boldsymbol{d} \,|\, q,M_{\rm chirp},\boldsymbol{\Lambda}\right)$, where $q\coloneqq M_{2}/M_{1}$, and $M_{\rm chirp}$ is a symmetric combination of $M_{1}$ and $M_{2}$. Here, $q\in\mathbb{R}^{+}$ if no ordering of masses is enforced, in which case the nuisance-marginalized likelihood for $q=Q$ is equal to that for $q=1/Q$ under exchange of properties. On the other hand, $0<q\leq1$ if $M_{1}$ is associated with the more massive component, which is the definition we choose.

The joint posterior distribution of interior parameters (omitting the global model as a conditional argument) is \explain{$\boldsymbol{y}$ is same as $\boldsymbol{\theta}$ in main text}
\begin{equation}
p(\boldsymbol{\theta},\boldsymbol{\varepsilon} \,|\,\boldsymbol{d})
\propto
p(\boldsymbol{d} \,|\, \boldsymbol{\theta},\boldsymbol{\varepsilon})p(\boldsymbol{\varepsilon}\,|\,\boldsymbol{\theta})p(\boldsymbol{\theta}),
\end{equation}
where $\boldsymbol{\theta}$ are EOS parameters, and $\boldsymbol{\varepsilon}$ are central (energy) densities, such that $(\boldsymbol{\theta},\varepsilon_{i})\mapsto(M_{i}, \Lambda_{i})$. The EOS parameters also operate as hyperparameters, in the simplest mode as an upper bound on the prior support of $\varepsilon\in[a,b(\boldsymbol{\theta})]$. The population-level prior density of binaries $p(\boldsymbol{\varepsilon}\,|\,\boldsymbol{\theta})$ is \textit{assumed} to be separable when one density is associated with a particular physical component:
\begin{equation}
p(\boldsymbol{\varepsilon}\,|\,\boldsymbol{\theta})
=
\mathop{\prod}_{i}
p(\varepsilon_{i}\,|\,\boldsymbol{\theta}),
\end{equation}
where $p(\varepsilon_{i}\,|\,\boldsymbol{\theta})$ is identical $\forall i$. Let us assume $f(\epsilon_{i})\sim U(a, b)$ for $b=b(\boldsymbol{\theta})$, where $f(\varepsilon_{i})$ is a monotone transformation such as a logarithm, which is our choice in the main text.

We decided to define one density parameter as that of the binary component with the highest central density (and thus total mass), and therefore $f(\varepsilon_{1})\geq f(\varepsilon_{2})$. Transforming the joint prior density function above, assuming $f(\epsilon_{i})=\epsilon_{i}$ for notational simplicity, yields
\begin{equation}
p(\varepsilon_{1},\varepsilon_{2}\,|\,\boldsymbol{\theta})
=
p(\varepsilon_{2}\,|\,\varepsilon_{1},\boldsymbol{\theta})p(\varepsilon_{1}\,|\,\boldsymbol{\theta})
\end{equation}
where
\begin{equation}
p(\varepsilon_{1}\,|\,\boldsymbol{\theta})=\frac{2(\varepsilon_{1} - a)}{(b-a)^{2}}
\quad\text{and}\quad
p(\varepsilon_{2}\,|\,\varepsilon_{1})
=
\frac{1}{(\varepsilon_{1} - a)},
\end{equation}
for support $\varepsilon_{1}\in[a,b]$ and $\varepsilon_{2}\in[a,\varepsilon_{1}]$. We then require (according to standard prior implementation for nested sampling)
\begin{equation}
x_{1}(\varepsilon_{1};\boldsymbol{\theta})
=
\mathop{\int}_{a}^{\varepsilon_{1}}
\frac{2(t - a)}{(b - a)^{2}}
dt
=
\frac{(\varepsilon_{1} - a)^{2}}{(b - a)^{2}};
\end{equation}
inverting yields
\begin{equation}
\varepsilon_{1}(x_{1};\boldsymbol{\theta})
=
a + (b - a)\sqrt{x_{1}}.
\end{equation}
Further,
\begin{equation}
x_{2}(\varepsilon_{2};\varepsilon_{1})
=
\frac{1}{(\varepsilon_{1} - a)}
\mathop{\int}_{a}^{t}
dt
=
\frac{\varepsilon_{2} - a}{(\varepsilon_{1} - a)}
\implies
\varepsilon_{2}(x_{2};x_{1}) = a + (\varepsilon_{1} - a)x_{2}.
\end{equation}

Returning to the likelihood function $p\left(\boldsymbol{d} \,|\, \boldsymbol{M},\boldsymbol{\Lambda}\right)$, the function is almost degenerate, with support (defined in terms of some small threshold value of the likelihood normalized to the global maximum) only for $|g(\boldsymbol{M})|\leq\epsilon$, where $g(\boldsymbol{M})$ is the chirp combination minus some (now) constant $M_{\rm chirp}$ and $\epsilon\in\mathbb{R}^{+}$ is small. In the limit $\epsilon\to0$, the likelihood function is degenerate in the $\boldsymbol{M}$-subspace. Generally, the marginal posterior density is then
\begin{equation}
p(\boldsymbol{\theta},\varepsilon_{1} \,|\, \boldsymbol{d})
\propto
p(\varepsilon_{1}\,|\,\boldsymbol{\theta})p(\boldsymbol{\theta})
\mathop{\int}
p(\boldsymbol{d}\,|\,\boldsymbol{M},\boldsymbol{\Lambda})
p(\varepsilon_{2}\,|\,\boldsymbol{\theta})
d\varepsilon_{2}.
\end{equation}
If $\epsilon\to0$, then
\begin{equation}
\begin{aligned}
p(\boldsymbol{\theta},\varepsilon_{1} \,|\, \boldsymbol{d})
&\propto
p(\varepsilon_{1}\,|\,\boldsymbol{\theta})p(\boldsymbol{\theta})
p(\boldsymbol{d}\,|\,M_{1},\boldsymbol{\Lambda})
\mathop{\int}
\delta(\varepsilon_{2} - h(\varepsilon_{1},\boldsymbol{\theta};M_{\rm chirp}))
p(\varepsilon_{2}\,|\,\boldsymbol{\theta})
d\varepsilon_{2}\\
&=
p(\varepsilon_{1}\,|\,\boldsymbol{\theta})p(\boldsymbol{\theta})
p(\boldsymbol{d}\,|\,M_{1},\boldsymbol{\Lambda})
\underbrace{p(h(\varepsilon_{1},\boldsymbol{\theta};M_{\rm chirp})\,|\,\boldsymbol{\theta})}_{\text{support}\;\varepsilon_{2}\in[a,b]},
\end{aligned}
\end{equation}
where $M_{2}$ is the total mass yielded via inversion of the chirp combination of $\boldsymbol{M}$, and $h(\varepsilon_{1}, \boldsymbol{\theta}; M_{\rm chirp})$ is the associated value of $\varepsilon_{1}$ for the EOS specified by $\boldsymbol{\theta}$. \explain{the dependence of variable $h(\ldots)$ on the EOS is now included.} Further, $\Lambda_{2}=\Lambda_{2}(\varepsilon_{1},\boldsymbol{\theta};M_{\rm chirp})$.
The last factor implies that we need to take the prior support of $\varepsilon_{2}$ into account given hyperparameters $\boldsymbol{\theta}$; note that if $h(\varepsilon_{1}, \boldsymbol{\theta}; M_{\rm chirp})$ has no solution, it will by definition not be within the prior support, and note also that the set of solutions to the inversion problem is generally a proper superset of the prior support.

However, because we enforced an order on the total masses, such that $M_{1}\geq M_{2}$, then
\begin{equation}
p(\boldsymbol{\theta},\varepsilon_{1} \,|\, \boldsymbol{d})
\propto
p(\varepsilon_{1}\,|\,\boldsymbol{\theta})p(\boldsymbol{\theta})
\mathop{\int}
p(\boldsymbol{d}\,|\,\boldsymbol{M},\boldsymbol{\Lambda})
p(\varepsilon_{2}\,|\,\varepsilon_{1},\boldsymbol{\theta})
d\varepsilon_{2}.
\end{equation}
If $\epsilon\to0$, then
\begin{equation}
\begin{aligned}
p(\boldsymbol{\theta},\varepsilon_{1} \,|\, \boldsymbol{d})
&\propto
p(\varepsilon_{1}\,|\,\boldsymbol{\theta})p(\boldsymbol{\theta})
p(\boldsymbol{d}\,|\,M_{1},\boldsymbol{\Lambda})
\mathop{\int}
\delta(\varepsilon_{2} - h(\varepsilon_{1},\boldsymbol{\theta};M_{\rm chirp}))
p(\varepsilon_{2}\,|\,\varepsilon_{1},\boldsymbol{\theta})
d\varepsilon_{2}\\
&=
p(\varepsilon_{1}\,|\,\boldsymbol{\theta})p(\boldsymbol{\theta})
p(\boldsymbol{d}\,|\,M_{1},\boldsymbol{\Lambda})
\underbrace{p(h(\varepsilon_{1},\boldsymbol{\theta};M_{\rm chirp})\,|\,\varepsilon_{1},\boldsymbol{\theta})}_{\text{support}\;\varepsilon_{2}\in[a,\varepsilon_{1}]}.
\end{aligned}
\label{eqn:delta-approximated GW likelihood}
\end{equation}
This means that if the most massive binary component is such that $h(\varepsilon_{1},\boldsymbol{\theta};M_{\rm chirp})\geq\varepsilon_{1}$, then the last factor imposes zero (approximate) posterior density at point $[\boldsymbol{\theta},h(\varepsilon_{1},\boldsymbol{\theta};M_{\rm chirp})]^{\top}$. Note that if the hyperparameters $\boldsymbol{\theta}$ merely delimit stability against radial perturbations, then $p(h(\varepsilon_{1},\boldsymbol{\theta};M_{\rm chirp})\,|\,\varepsilon_{1},\boldsymbol{\theta})$ loses dependence, becoming $p(h(\varepsilon_{1},\boldsymbol{\theta};M_{\rm chirp})\,|\,\varepsilon_{1})$. When transforming $M_1$ in Equation \ref{eqn:delta-approximated GW likelihood} to the mass ratio $q$ we obtain the GW-related part of Equation \ref{eq:eq3} in the main text.

In reality the likelihood function is finite for $|g(\boldsymbol{M})|\leq\epsilon$ with $\epsilon$ finite, with the marginal posterior distribution of the chirp combination having 90\% credible interval about the median of $M_{\rm chirp}=1.186_{-0.001}^{+0.001}$~M$_\odot$. If we are to account for this likelihood information accurately, but avoid sampling from a relatively very diffuse prior in the central densities and thus in the chirp mass, then we need to perform fast marginalization over the central density $\varepsilon_{2}$ to decrement the dimensionality of the sampling space. To proceed, we need to examine the \textit{conditional} likelihood function
\begin{equation}
L(\varepsilon_{2};\varepsilon_{1},\boldsymbol{\theta})
=
p(\boldsymbol{d} \,|\, \boldsymbol{M}, \boldsymbol{\Lambda}).
\end{equation}
Given $M_{1}$, there is one maximum in the conditional likelihood function when slicing through the chirp-sensitive likelihood function. Further, $M_{2}=M_{2}(\varepsilon_{2};\boldsymbol{\theta})$, whilst not a monotone transformation in the presence of unstable branches, is ordered, meaning there should exist one maximum with respect to $\varepsilon_{2}$. In order to marginalize we therefore aim to first approximate the central density $\varepsilon_{2}$ that maximizes the conditional likelihood function. We thus approximate---via inversion---$h(\varepsilon_{1},\boldsymbol{\theta};M_{\rm chirp}^{\prime})$, as above, where $M_{\rm chirp}^{\prime}$ is now the median chirp mass in the marginal posterior distribution of $M_{\rm chirp}$. Given this estimator of the maximum, we can generate bounds for numerical integration. For instance, we can integrate on the interval $\varepsilon_{2}\in[\alpha,\beta]$, where $\alpha\coloneqq h(\varepsilon_{1},\boldsymbol{\theta};M_{\rm chirp}^{\prime}-n\mathcal{C}_{l})$ and $\beta\coloneqq h(\varepsilon_{1},\boldsymbol{\theta};M_{\rm chirp}^{\prime}+n\mathcal{C}_{u})$ where $n$ is some integer and the X\% marginal credible interval on the chirp mass is CI$_{\rm X\%}=\{M_{\rm chirp}\colon M_{\rm chirp}\in[M_{\rm chirp}^{\prime}-\mathcal{C}_{l}, M_{\rm chirp}^{\prime}+\mathcal{C}_{u}]\}$. We can then perform numerical quadrature where the integrand requires numerical integration to transform $(\varepsilon_{2},\boldsymbol{\theta})\mapsto(M_{2},\Lambda_{2})$. If an order is imposed on the central densities, then the upper-bound for quadrature is $\text{min}[h(\varepsilon_{1},\boldsymbol{\theta};M_{\rm chirp}^{\prime}+n\mathcal{C}_{u}), \varepsilon_{1}]$.

Another approach would be to inject likelihood information into the \textit{prior} density function of the mass $M_{2}$ (and thus of the central density $\varepsilon_{2}$). The prior \textit{support} of $M_{2}$ (and thus $\varepsilon_{2}$) is restricted to some narrow interval corresponding to a narrow interval in chirp mass about the marginal median value. This is uncomfortable to have to rely on, one reason being that prior predictive probabilities may be compromised for model comparison; another is that a prior may be defined on the space of $M_{2}$ that is inconsistent with the prior density $p(\varepsilon_{1}\,|\,\boldsymbol{\theta})$, or a prior is defined directly on the space of the chirp mass, leading to a similar inconsistency. Given this modification of the prior support relative to the protocol outlined above, (nested) sampling proceeds without decrementing the dimensionality, but at far lower cost and effectively without risk of error due to insufficient resolution.

In this work we compared two approaches quantitatively: (i) the delta-function approximation---specifically Equation~(\ref{eqn:delta-approximated GW likelihood}); and (ii) the approach wherein likelihood information is injected into the prior via modification of the support, in the form of a narrow prior on the chirp mass about the median in marginal posterior mass; see the text following Equation~(\ref{eq:eq3}). The joint posterior distribution of the EOS parameters is consistent for both approaches.

\begin{figure}
    \centering
    \includegraphics[width=.48\textwidth]{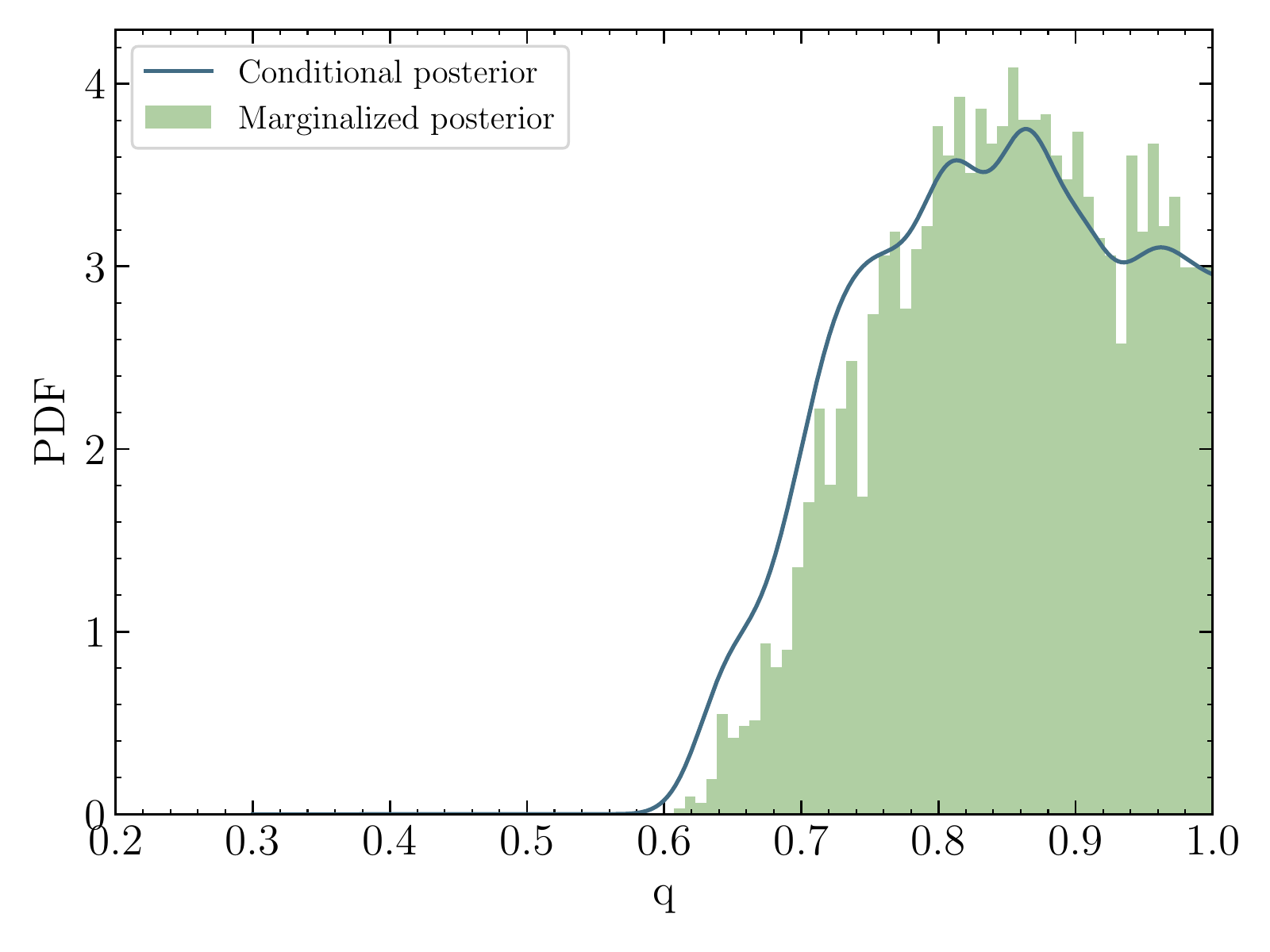}
    \includegraphics[width=.48\textwidth]{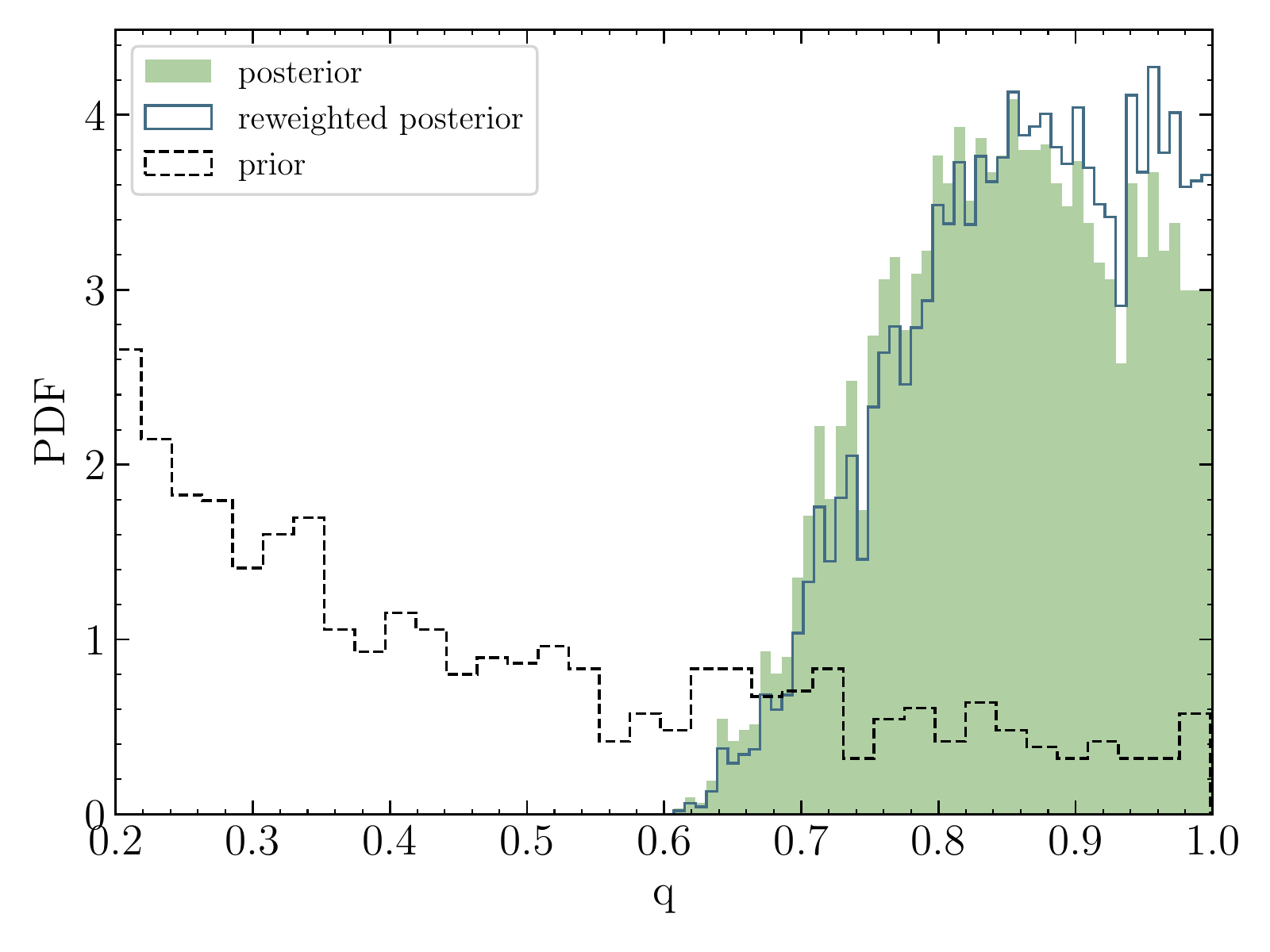}
    \caption{Left panel: comparison of the distribution of the mass ratio $q$ when we use the posterior distribution marginalized over the chirp mass (green) or the posterior distribution conditional on a given chirp mass (blue). Both distributions are very similar, which is why we argue that the usage of the marginalized distribution throughout the Letter is justified. Right panel: prior distribution on the mass ratio $q$ in the black dashed lines, the posterior distribution on $q$ in green, and the reweighted posterior that corresponds to a flat prior in $q$ in blue. Note that the prior distribution on $q$ is only for a small interval in chirp mass where there is posterior support, which is why there is more support toward lower-than higher-mass ratios. We conclude that there is slightly more posterior support for equal mass ratios when reweighting but this is negligible compared to the statistical uncertainty in GW170817.}
    \label{fig:fig7}
\end{figure}

\section{Approximations to the Likelihood Function}
\label{AppendixB}
\subsection{The NS-NS Scenario}
The posterior distribution on the two binary components from GW170817 derived by the LVC \citep[see][]{GW170817sourceproperties} is related to the nuisance-marginalized likelihood function through
\begin{equation}
    p(\Lambda_1, \Lambda_2, M_1, M_2 ~|~ \bm{d}_{\rm{GW}}) \propto p(\Lambda_1, \Lambda_2, M_1, M_2) ~ p(\bm{d}_{\rm{GW}} ~|~ \Lambda_1, \Lambda_2, M_1, M_2),
\end{equation}
where we have only shown the four parameters of interest here. In this work we have equated the two by arguing that the prior distribution $p(\Lambda_1, \Lambda_2, M_1, M_2)$ is jointly flat. Furthermore we have performed a coordinate transformation from the two component masses to the mass ratio $q$ and the chirp mass $M_{\rm{chirp}}$ to get
\begin{equation}
 p(\Lambda_1, \Lambda_2, q, M_{\rm{chirp}} ~|~ \bm{d}_{\rm{GW}}).
\end{equation}
As the chirp mass is constrained to a very small range we fix its value to $M_{\rm{chirp}} = 1.186$ \msol such that we can write a conditional distribution
\begin{equation}
    p(\Lambda_1, \Lambda_2, q ~|~ \bm{d}_{\rm{GW}}, M_{\rm{chirp}}),
\end{equation}
which is a slice through the posterior distribution. We use, however, the marginalized posterior distribution $p(\Lambda_1, \Lambda_2, q ~|~ \bm{d}_{\rm{GW}} )$ in this Letter, which we compare to the conditional distribution in the left panel of Figure \ref{fig:fig7}.

We furthermore note that when transforming from component masses to mass ratio and chirp mass the prior on these quantities is no longer jointly flat. We therefore check that the posterior distribution on $q$ does not change significantly when applying a weighting to the distribution to ensure a jointly flat prior. We show the difference between the distributions in the right panel of Fig. \ref{fig:fig7}. 

\subsection{The NS-BH Scenario}
\begin{figure}
    \centering
    \includegraphics[width=.48\textwidth]{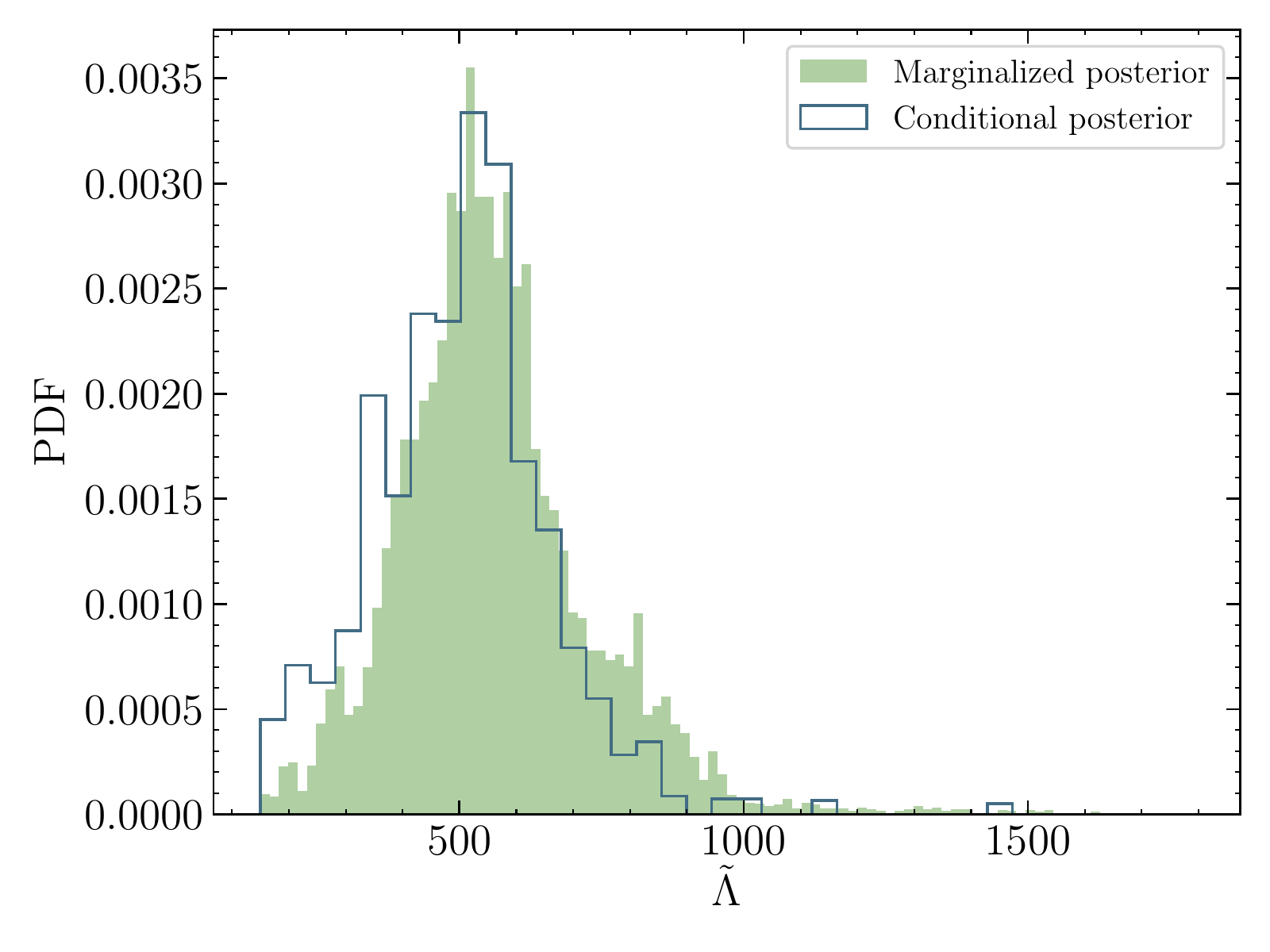}
    \caption{Comparison between the posterior distribution on $\Tilde{\Lambda}$ when marginalizing over $\Lambda_1$ and when only considering values of $\Lambda_1 < 30$. The range $\Lambda_1<30$ is chosen in order to well-approximate the conditional distribution at $\Lambda_1 = 0$ whilst reducing noise in the histogram. The distributions shown here have already implemented zero support on parameter sets that result in $M_{\rm{rem}}<0.1$ \msol and uniform support where $M_{\rm{rem}}>0.1$ based on the electromagnetic analysis in Section \ref{sec:BHNS} and \citet{Hinderer19}. The difference between the two distributions is small enough to justify the approximation made in this Letter.}
    \label{fig:fig8}
\end{figure}

Finally we have approximated the likelihood function for the NS-BH case by transforming $\Lambda_1$ and $\Lambda_2$ to the parameter $\Tilde{\Lambda}$ and marginalizing over $\Lambda_1$, which is different from taking a slice through the posterior at $\Lambda_1 = 0$. However, to compute the constraints from the EM counterpart the latter is not possible since there are not enough samples in the posterior that have $\Lambda_1 = 0$. As an alternative we approximate $\Lambda_1 = 0$ by taking the samples in the distribution where $\Lambda_1 < 30$, such that we are not dominated by stochastic noise of having too few samples. In Fig. \ref{fig:fig8} we compare the two approaches when we set the posterior support for parameter sets that result in $M_{\rm{rem}}<0.1$ \msol to zero, while setting uniform support everywhere else. Including these constraints from the EM counterpart, we conclude that there is a negligible change in the posterior distribution of $\Tilde{\Lambda}$.   

\end{document}